\documentclass[useAMS,usenatbib]{mn2e}

\usepackage{times}

\usepackage{amssymb}
\usepackage{psfig}

\newcommand{\etal}{{\it et al.}\ }

\title[Broad-band colors and photometric properties of galaxy models]
{Broad-band colors and overall photometric properties of template galaxy models from
stellar population synthesis}
\author[A. Buzzoni]{Alberto Buzzoni\\
INAF - Osservatorio Astronomico di Bologna, Via Ranzani 1 - 40127 Bologna, Italy\\
{\sf e-mail: buzzoni@bo.astro.it}}

\begin{document}

\date{Accepted ... Received; in original form }

\pagerange{\pageref{firstpage}--\pageref{lastpage}} \pubyear{2005}

\maketitle

\label{firstpage}

\begin{abstract}
We present here a new set of evolutionary population synthesis models 
for template galaxies along the Hubble morphological sequence. The models, that account
for the individual evolution of the bulge, disk, and halo components,
provide basic morphological features, along with bolometric luminosity and color evolution 
(including Johnson/Cousins, Gunn $g,r,i$, and Washington $C,M,T_1,T_2$ photometric systems) 
between 1 and 15 Gyr. Luminosity contribution from residual gas is also evaluated,
both in terms of nebular continuum and Balmer-line enhancement.

Our theoretical framework relies on the observed colors of present-day 
galaxies, coupled with a minimal set of physical assumptions related to SSP evolution theory, 
to constrain the overall distinctive properties of galaxies at earlier epochs.
A comparison with more elaborated photometric models, and with empirical sets of reference
SED for early- and late-type galaxies is accomplished, in order to test output reliability and 
investigate internal uncertainty of the models.

The match with observed colors of present-day galaxies tightly constrain 
the stellar birthrate, $b$, that smoothly increases from E to Im types.
The comparison with observed SN rate in low-redshift galaxies shows, as well, a pretty good
agreement, and allows us to tune up the inferred star formation activity and the SN and Hypernova rates
along the different galaxy morphological types. Among others, these results could find useful 
application also to cosmological studies, given for instance the claimed relationship between 
Hypernova events and Gamma-ray bursts.

One outstanding feature of model back-in-time evolution is the 
prevailing luminosity contribution of the bulge at early epochs. 
As a consequence, the current morphological look of galaxies might  drastically change when
moving to larger distances, and we discuss here how sensibly this bias could affect
the observation (and the interpretation) of high-redshift surveys.

In addition to broad-band colors, the modeling of Balmer line emission
in disk-dominated systems shows that striking emission 
lines, like $H\alpha$, can very effectively track stellar birthrate in a galaxy. For these features to
be useful age tracers as well, however, one should first assess the real change of $b$ vs.\ time
on the basis of supplementary (and physically independent) arguments.
\end{abstract}


\section{Introduction}

Since its early applications to extragalactic studies \citep{st71,tg78}, 
stellar population synthesis has been the natural 
tool to probe galaxy evolution. The works of \citet{searle} and \citet{larson} 
provided, in this sense, a first important reference for a unitary 
assessment of spectrophotometric properties of early- and late-type systems 
in the nearby Universe, while the contribution of \citet{bk80}, \citet{ll84},
\citet{yoshii}, among others, represent 
a pionering attempt to extend the synthesis approach also to unresolved 
galaxies at cosmological distances.

In this framework, color distribution along the Hubble sequence has readily been recognized as 
the most direct tracer for galaxy diagnostics; a tight relationship exists in fact 
between integrated colors and morphological type, through the relative 
contribution of bulge and disk stellar populations \citep{koppen,aj91}. 
Ongoing star formation, in particular, is a key mechanism to modulate galaxy 
colors, especially at short wavelength \citep{lt78,k98}, while visual and infrared luminosity are 
better sensitive to the global star formation history \citep{quirk,sandage86,gs96}. 
External environment conditions could also play a role, as 
well as possible interactions of galaxies with embedding giant haloes, like in 
some CDM schemes \citep{ft92,ft94,pf94}.

In this work I want to try a simple heuristic approach to galaxy photometric 
evolution relying on a new family of theoretical template models to account 
for the whole Hubble morphological sequence.
Galaxy evolution is tracked here in terms of the individual history of the 
composing sub-systems, including the bulge, disk and halo;
the present discussion completes the analysis already undertaken in an accompanying paper 
\citep[][hereafter Paper I]{b02}, and relies on the evolutionary population 
synthesis code developed previously \citep[][hereafter B89 and B95, respectively]{b89,b95}. 

A main concern of this work is to provide the user with a quick reference tool
to derive broad-band colors and main morphological parameters of galaxies 
allover their ``late'' evolutionary stages (i.e.\ for $t \ga 1$~Gyr).
This should be the case for most of $z \la 3$ systems, for which the 
formation event is mostly over and the morphological design already in place. 

Within minor refinements, the present set of models has already been successfully 
used by \citet{massarotti} in their photometric study of the Hubble 
Deep Field galaxies; a further application of these templates is also due to
\citet{b05}, to assess the evolutionary properties of the planetary 
nebula population in bright galaxies and the intracluster medium.
Compared to other more ``physical'' (and entangled) 
approaches, I believe that a major advantage of this simplified treatment 
is to allow the user maintaining a better control of the theoretical output, 
and get a direct feeling of the changes in model properties as far as one or more 
of the leading assumptions are modified.

Models will especially deal with the stellar component, which is obviously 
the prevailing contributor to galaxy luminosity. Residual gas acts more 
selectively on the integrated spectral energy distribution (SED) by enhancing 
monochromatic emission, like for the Balmer lines. 
As far as galaxy broad-band colors are concerned, in the present age range, its 
influence is negligible and can be treated apart in our discussion.
Internal dust could play, on the contrary,  a more important role,
especially at short wavelength ($\lambda \la 3000$~\AA). Its impact
for high-redshift observations has been discussed in some detail in Paper I.

In this paper we will first analyze, in Sec.~2, the basic components of the 
synthesis model, taking the Milky Way as a main reference to tune up some
relevant physical parameters for spiral galaxies. A set of color fitting functions is also
given in this section, in order to provide the basic analytical tool to
compute galaxy luminosity for different star formation histories.

Model setup is considered in Sec.~3, especially dealing with the
disk physical properties; metallicity and stellar birthrate will be constrained 
by comparing with observations and other theoretical studies.
In Sec.~4 we will assemble our template models, providing colors and other
distinctive features for each galaxy morphological type along the
Hubble sequence, from E to Im. A general sketch of back-in-time evolution
is outlined in this section, focussing on a few relevant aspects that deal
with the interpretation of high-redshift data. 
Section~5 discusses the contribution of the residual gas; we will evaluate 
here the nebular luminosity and derive Balmer emission-line evolution. The main issues 
of our analysis are finally summarized in Sec.~6.

\section{Operational tools}

Our models consist of three main building blocks: we will consider a central
bulge, a disk, and an external halo.
It is useful to track evolution of each block individually, in terms of 
composing simple stellar populations (SSPs), taking advantage of the powerful 
theoretical formalization by \citet{tinsley} and \citet{rb86}.

If SSP evolution is known and a star formation rate (SFR) can be assumed 
vs.\ time, the general relation for integrated luminosity of a stellar system is
\begin{equation}
L_{\rm gal}(t)  = \int_{0}^t L_{\rm SSP}(\tau)\ SFR(t-\tau)\ d\tau.
\label{eq:lgal}
\end{equation}
Operationally, the integral in eq.~(\ref{eq:lgal}) is computed by discrete time 
steps, $\Delta \tau$, taking the lifetime of the most massive stars in the initial 
mass function (IMF),  $t_{\rm min}$, as a reference,
so that $\Delta \tau = t_{\rm min}$.

In a closed-box evolution, galaxy SFR is expected to be a decreasing function 
of time \citep[e.g.][]{tinsley, ay86}; on the other hand, if 
fresh gas is supplied from the external environment, then an opposite trend 
might even be envisaged.
One straighforward way to account for this wide range of evolutionary paths 
is to assume a power law such as SFR~$= K\ t^{-\eta}$, 
with $\eta < 1$.\footnote{In our notation it must always be $t \ge t_{\rm min}$ as,
in force of eq.~(\ref{eq:lgal}), gas consumption proceeds over 
discrete time steps corresponding to the lifetime of high-mass stars.}

As we pointed out in Paper I, an interesting feature of this simple 
parameterization is that stellar birthrate,
\begin{equation}
b = {{{\rm SFR}(t)}\over {\langle{\rm SFR}\rangle}} = (1-\eta),
\label{eq:birthrate}
\end{equation}
is a time-independent function of SFR and becomes therefore an intrinsic 
distinctive parameter of the galaxy model.\footnote{As SFR must
balance the net rate of change of the residual gas [i.e.\ ${\rm SFR}(t) = 
-\dot{g}(t)$], then ${\rm SFR}(t) = (\Delta g / t)\ (1-\eta)$
and $\langle {\rm SFR} \rangle = t^{-1}\int {\rm SFR}\,d\tau = \Delta g / t$. 
Stellar birthrate $b = (1-\eta)$ 
can therefore be regarded as an ``efficiency factor'' in the gas-to-star conversion.}

\subsection {SSP evolution}

In order to properly apply eq.~(\ref{eq:lgal}) to the different evolutionary 
cases, we need first to secure its basic ``ingredient'' by modeling SSP 
luminosity evolution.
The original set of B89 and B95 population synthesis models, and its following 
upgrade and extension as in Paper I, especially dealt with evolution of low- and 
intermediate-mass stars, with $M \la 2~M_\odot$ and main sequence (MS) 
lifetime typically greater than 1 Gyr. As a striking feature in this mass 
range, red giant branch sets on in stars with a degenerate Helium core, 
tipping at high luminosity ($\log L/L_\odot \sim 3.3$) with the so-called 
Helium-flash event \citet{sg78}. 

\begin{figure*}
\centerline{
\psfig{file=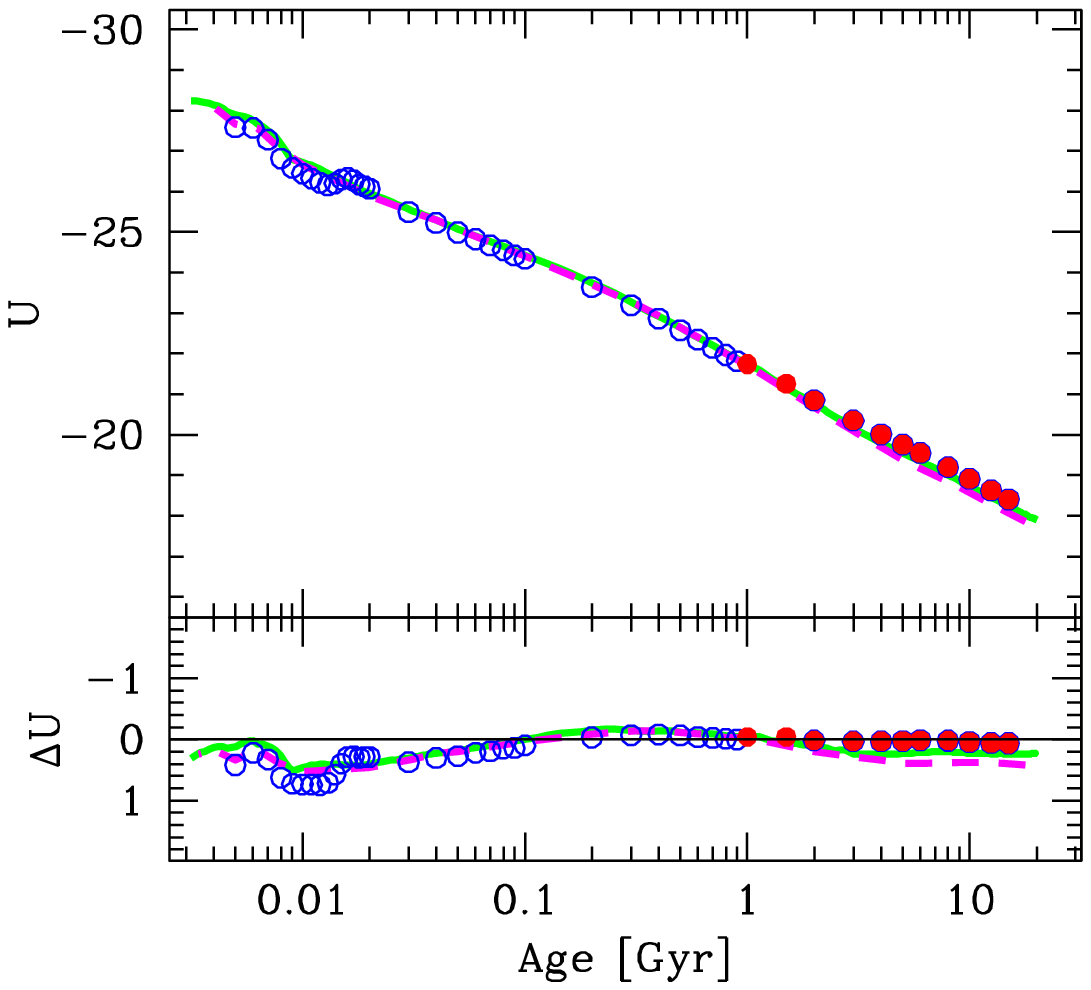,width=0.39\hsize,clip=}
\psfig{file=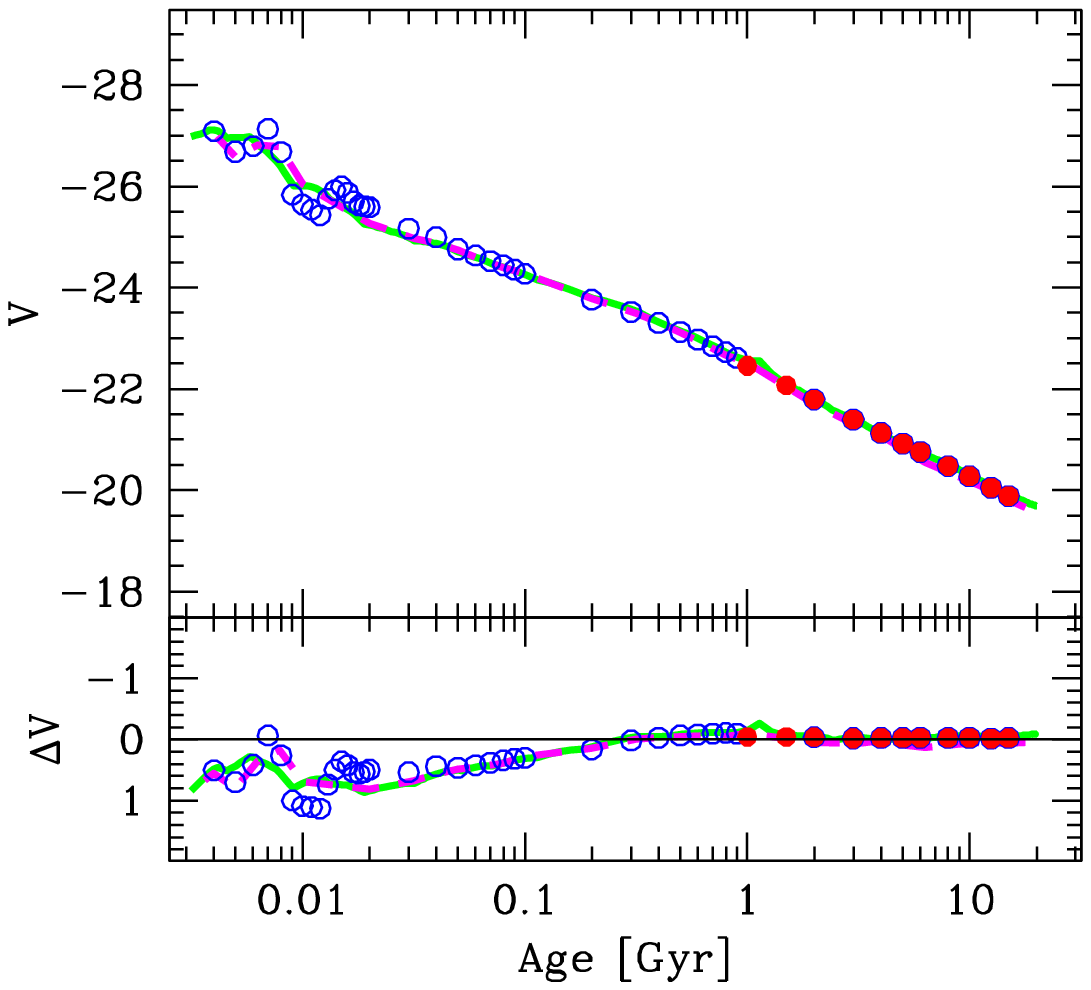,width=0.39\hsize,clip=}}

\centerline{
\psfig{file=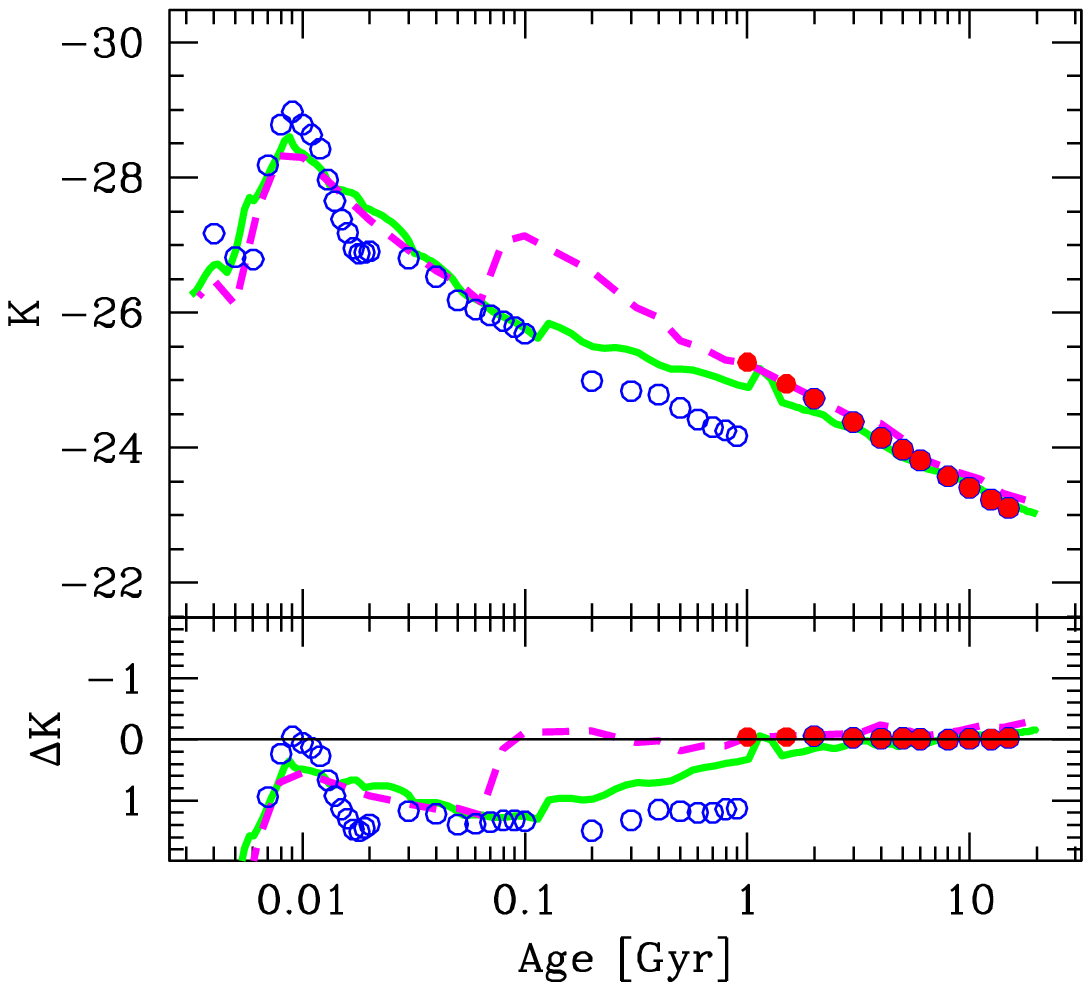,width=0.39\hsize,clip=}
\psfig{file=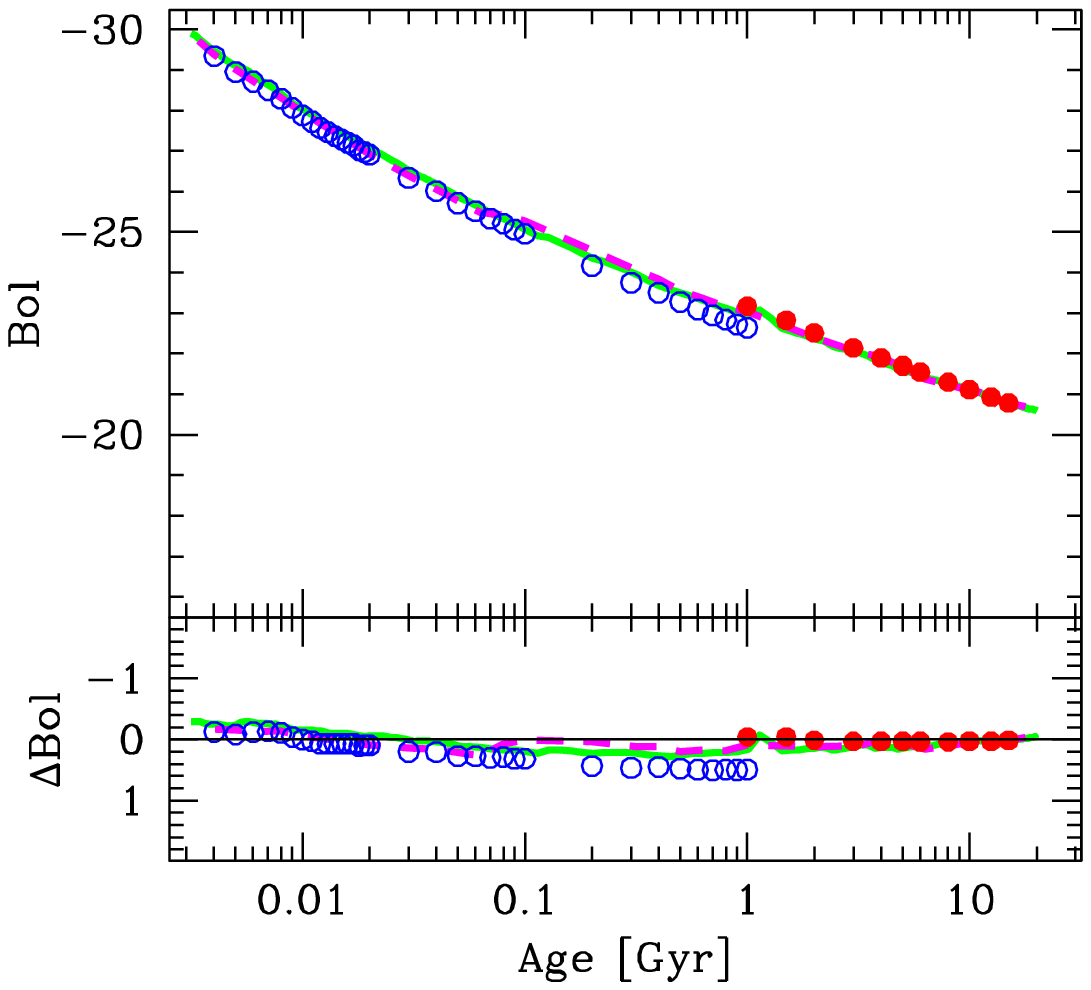,width=0.39\hsize,clip=}}
\caption{
{\it Upper plots of each panel}: luminosity evolution of present SSP models 
(solid dots), for solar metallicity and Salpeter IMF, is compared with other 
theoretical outputs according to \citet[][open dots]{leitherer}, \citet[][dashed line]{bcf94}, 
and \citet[][solid line]{bc03}.
Total mass is scaled to $M_{\rm SSP} = 10^{11}~M_\odot$ throughout, with stars
in the range between 0.1--120~$M_\odot$. The \citet{leitherer} model has 
been slightly increased in luminosity (by about 0.06 mag in bolometric, at 
1 Gyr) to account for the missing luminosity contribution from low-MS stars
($M_* \leq 1.0~M_\odot$).
\protect \\
{\it Lower plots of each panel}: model residuals with respect to the fitting 
functions of Table~\ref{coeff}. See text for discussion.}
\label{fig_all}
\end{figure*}

For younger ages (i.e.\ $t \la 1$~Gyr), evolution is less univocally 
constrained, as convection and mass loss via stellar winds (both depending 
on metallicity) sensibly modulate the evolutionary path of high-mass stars 
across the H-R diagram on timescales as short as $\sim 10^6$~yr \citep{deloore,mc94}. 
Such a quicker Post-MS evolution could also lead to a biased sampling of stars 
in the different phases, even at galaxian mass-scale, inducing intrinsic 
uncertainties in the SSP spectrophotometric properties independently of 
the input physics details (see \citealp{cervino02} and \citealp{cvg03}, for a brilliant 
assessment of this problem).

A combined comparison of different SSP models, over the full range of stellar 
masses, is displayed in Fig.~\ref{fig_all}, where we matched our SSP model
sequence for solar metallicity and \citet{sal} IMF with the corresponding 
theoretical output from \citet{leitherer}, \citet{bcf94} and 
\citet[][their ``Padova 1994'' isochrone setup]{bc03}.
Luminosity in the Johnson $U$, $V$, $K$ bands, and in bolometric as well, 
is explored in the four panels of the figure, referring to evolution of a 
$10^{11}~M_\odot$ SSP, with stars in the range $0.1 \leq M_*/M_\odot \leq 120$.
To consistently compare with the other outputs, the Leitherer \etal models 
have been slightly increased in luminosity (by some $\Delta {\rm mag} = -0.06$ 
in bolometric at 1 Gyr) to account for the missing low-MS ($M_* < 1~M_\odot$)
contribution, according to the B89 estimates.

Figure~\ref{fig_all} shows a remarkable agreement between the four theoretical
codes; in particular, ultraviolet and bolometric evolution is fairly well 
tracked over nearly four orders of magnitude of SSP age. The lack of AGB 
evolution in the \citet{leitherer} code is, however, especially evident in 
the $K$  plot, where a glitch of about 1 mag appears at the match with our 
model sequence about $t \sim 1$~Gyr.
To a lesser extent, also the $K$-band contribution of red giant stars seems to be
partly undersized in the \citet{bc03} models between $10^8$ and $10^9$~yrs, 
probably be due to interpolation effects across stellar tracks
in the relevant range of mass (i.e.\ $M_* = 5 \to 2~M_\odot$).

\setcounter{table}{0}

\begin{table*}
\begin{minipage}{153mm}
\caption{SSP magnitude fitting functions$^{(a)}$}
\label{coeff}
\begin{tabular}{lrccccclcl}
\hline
Band & $\lambda_{\it eff}$ & $\log f_o^{(b)}$ & \multicolumn{5}{c}{mag = $(\alpha'+\alpha''{\rm [Fe/H]}) \log t_9 + \beta {\rm [Fe/H]} +\gamma +\delta({\rm t}_9,{\rm [Fe/H]})$} & $\sigma$ & System$\qquad$ \\
     &                     &                  & \multicolumn{5}{c} {\hrulefill} &   &   \\
     & [\AA]              & [erg s$^{-1}$ cm$^{-2}$ \AA$^{-1}$] & $\alpha'$ & $\alpha''$ & $\beta$ & $\gamma$ & $\delta({\rm t}_9,{\rm [Fe/H]})$ & [mag] &  \\
\hline
U     & 3650  & --8.392 & 2.743 & 0.26 & ~~0.006   &  ~\,2.872 & +0.09\,10$^{\rm [Fe/H]}$ & $\pm 0.026$ & Johnson \\
C     & 3920  & --8.275 & 2.560 & 0.16 & --0.024   &  ~\,2.853 & +0.09\,10$^{\rm [Fe/H]}$ & $\pm 0.022$ & Washington \\
B     & 4420  & --8.205 & 2.390 & 0.08 & ~~0.033   &  ~\,2.909 & &$\pm 0.020$ & Johnson \\
M     & 5060  & --8.351 & 2.232 & 0.08 & --0.060   &  ~\,2.477 & &$\pm 0.020$ & Washington \\
g     & 5170  & --8.384 & 2.214 & 0.08 & --0.069   &  ~\,2.401 & &$\pm 0.019$ & Gunn \\
V     & 5500  & --8.452 & 2.163 & 0.08 & --0.094   &  ~\,2.246 & &$\pm 0.019$ & Johnson \\
T$_1$ & 6310  & --8.632 & 2.072 & 0.08 & --0.160   &  ~\,1.835 & &$\pm 0.019$ & Washington \\
R$_C$ & 6470  & --8.670 & 2.048 & 0.08 & --0.176   &  ~\,1.756 & &$\pm 0.019$ & Cousins \\
r     & 6740  & --8.527 & 2.037 & 0.08 & --0.189   &  ~\,2.147 & &$\pm 0.019$ & Gunn \\
R     & 7170  & --8.790 & 2.006 & 0.08 & --0.207   &  ~\,1.546 & &$\pm 0.019$ & Johnson \\
I$_C$ & 7880  & --8.936 & 1.977 & 0.08 & --0.275   &  ~\,1.226 & &$\pm 0.019$ & Cousins \\
T$_2$ & 7940  & --8.938 & 1.972 & 0.08 & --0.255   &  ~\,1.250 & &$\pm 0.019$ & Washington \\
i     & 8070  & --8.653 & 1.969 & 0.08 & --0.259   &  ~\,1.981 & &$\pm 0.019$ & Gunn \\
I     & 9460  & --9.136 & 1.923 & 0.08 & --0.327   &  ~\,0.944 & &$\pm 0.020$ & Johnson \\
J     & 12500 & --9.526 & 1.863 & 0.08 & --0.444   &  ~\,0.333 & &$\pm 0.018$ & Johnson \\
H     & 16500 & --9.965 & 1.833 & 0.08 & --0.551   &  --0.375 & &$\pm 0.016$ & Johnson \\ 
K     & 22000 & --10.302~~ & 1.813 & 0.08 & --0.614   &  --0.561 & &$\pm 0.016$ & Johnson \\
      &       &            &       &      &           &          &   &           &   \\
Bol   &       &            & 1.923 &      & --0.324   & ~\,1.623 & $-0.1\,t_9^{-0.5}$ & $\pm 0.014$ &  \\
\hline
\end{tabular}
\medskip

$^{(a)}$ For a Salpeter IMF with stars in the mass range $0.1 \leq M/M_\odot \leq 120$.\\
$^{(b)}$ $\log f = -0.4\,{\rm mag} + \log f_o$. For the bolometric: $\log L/L_\odot = -0.4\,({\rm Bol} - 4.72)$.
\end{minipage}
\end{table*}

Definitely, SSP evolution appears to be best tracked by the isochrone-synthesis models of \citet{bcf94};
like in our code, these models meet the prescriptions of the so-called ``Fuel Consumption Theorem'' 
\citep{rb86}, and self-consistently account for the AGB energetic budget down to the onset 
of the SN\,II events (about $t \simeq 10^8$~yr). 
In any case, as far as early SSP evolution is concerned, a combined analysis of 
Fig.~\ref{fig_all} makes clear the intrinsic uncertainty of the synthesis
output in this particular age range, mainly as a result of operational and 
physical differences in the treatment of Post-MS evolution 
\citep[cf.][for further discussion on this subject]{charlot}.

\subsection {SSP fitting functions}

Model setup, like in case of composite stellar populations according to 
eq.~(\ref{eq:lgal}), would be greatly eased if we could manage the problem 
semi-analytically, in terms of a suitable set of SSP magnitude fitting 
functions. An important advantage in this regard is that also intermediate 
cases for age and/or metallicity could readily be accounted for in our 
calculations.

For this task we therefore considered the B89 and B95 original dataset of SSP 
models (and its further extension, as in Paper I), with Salpeter IMF and
red horizontal branch morphology (cf.\ B89 for details).
In addition to the original Johnson photometry, we also included here the 
Cousins ($R_C, I_C$), Gunn ($g,r,i$), and Washington ($C, M, T_1, T_2$) band 
systems. The works of \citet{bessell}, \citet{tg76}, \citet{schneider},
and \citet{canterna} have been referred to for the different system definition 
\citep[see also][for an equivalent calibration of the Washington 
colors based on the B89 models]{cf96}.
Our analysis will therefore span a wide wavelength range, from 3600 \AA\ 
($U$ band), to 2.2 $\mu m$ ($K$ band), sampling SSP luminosity at roughly 
$\Delta \lambda \sim 400$~\AA\ steps in the spectral window between 
3600--9000~\AA. 

For SSP magnitude evolution, a general fit was searched for in the form 
\begin{equation}
{\rm mag} = (\alpha'+\alpha''\,{\rm [Fe/H]})\ \log t_9 + \beta\ {\rm [Fe/H]} 
+ \gamma + \delta({\rm t}_9,{\rm [Fe/H]}),
\end{equation}
where $t_9$ is SSP age, expressed in Gyr. 
The whole set of the fitting coefficients is summarized in Table~\ref{coeff} 
(columns 4 to 8); note that the $U$ magnitude scale in the table already 
accounts for the corrections devised in B95. Column 2 of the table reports 
the effective wavelength for each photometric passband, as directly computed
from the adopted filter profile. 

Mass scaling for the SSP fitting functions in Table~\ref{coeff} can be done 
by adding to all magnitudes an offset 
\begin{equation}
\Delta {\rm mag} = -2.5 \log M_{\rm tot} +2.5 \log (M/L_{\rm V})_{15},
\end{equation}
where both SSP total mass and the V-band mass-to-light ratio (estimated 
at $t = 15$~Gyr) are in solar units. A quite accurate fit (within a 
$\pm 1.5$\% relative scatter) for the latter quantity over the metallicity 
range of our models is:
\begin{equation}
\left({M\over{L_{\rm V}}}\right)_{15} = 3.23\ ({\rm [Fe/H]} +1.5)^2 +6.41.
\label{eq:ml}
\end{equation}
Again, the fit holds for stars in the range 0.1--120~$M_\odot$ and a 
Salpeter IMF.

The whole set of fitting functions in Table~\ref{coeff} is optimized in the 
range [Fe/H]~$= [-1.5, +0.5]$. For $t \geq 1$ Gyr, model grid is fitted 
with an internal accuracy better than $\pm 0.02$ mag, slightly worsening in 
the ultraviolet, as reported in column 9 of the table.

The residual trend of the \citet{leitherer}, \citet{bcf94} and \citet{bc03} 
SSP models with respect to our fitting functions 
is displayed in the lower plot of each panel of Fig.~\ref{fig_all}.  Reference
equations give a fully adequate description of SSP luminosity evolution well 
beyond the nominal age limits of our model grid, and span the whole range of AGB 
evolution, for $t \geq 10^8$~yr. At early epochs, of course, fit predictions partially 
miss the drop in the SSP integrated luminosity, when core-collapsed evolution of 
high-mass stars ends up as a SN burst thus replacing the standard AGB phase
(we will return on this important feature in Sec.\ 4.1). As far as composite stellar 
populations are concerned, however, the induced uncertainty of fit extrapolation on 
the total luminosity of the system is much reduced since eq.~(\ref{eq:lgal}) averages 
SSP contribution over time.\footnote{For the illustrative 
case of a SFR constant in time, from our calculations we estimate that the effect of the 
``AGB glitch'' in the first $10^8$~yr of SSP evolution, with respect to a plain extrapolation 
of Table~1 fitting functions, reflects on the integrated colors of the composite 
stellar population by $\Delta (B-V) \simeq \Delta (V-K) \lesssim 0.06$~mag for 1~Gyr
models. In terms of absolute magnitude, the luminosity drop amounts to
a maximum of $\sim 0.25$~mag in the $K$ band, at 1~Gyr (and for $\eta = 0$),
reducing to a $\Delta K \simeq 0.1$~mag for 15~Gyr models.
All these figures will further reduce at shorter wavelength and when SFR decreases with 
time (i.e.\ for $\eta > 0$); they could be taken, therefore, as a conservative upper limit 
to the internal uncertainty of our models. In any case, in this work we 
will restrain our analysis only to galaxies older than 1~Gyr.}

\begin{figure}
\psfig{file=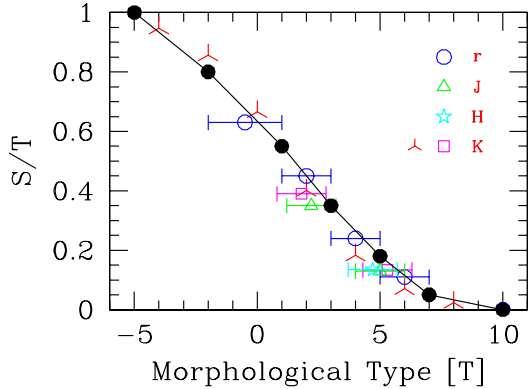,width=0.9\hsize,clip=}
\caption{The adopted red/infrared calibration for the S/T morphological 
parameter, defined as $L_{\rm spheroid}/L_{\rm tot}$ ($\bullet$ markers), 
as derived from the data of Kent (1985; $\circ$ markers), de Jong (1996; 
reversed ``Y'' markers), Giovanardi \& Hunt (1988; $\sq$, $\triangle$ 
and $\star$ markers for later-type spirals at $T \sim 5$), and 
Moriondo \etal (1998; $\triangle$ and $\sq$ markers for early-type spirals, 
about $T \sim 2$). Photometric bands of observations are labeled top right
in the plot.}
\label{st_calib}
\end{figure}

\section{Model setup: the basic building blocks}

To consistently assemble the three main building blocks of our synthesis models 
we mainly relied on the \citet{kent}  galaxy decomposition profiles, that probe 
the spheroid (i.e.\ bulge+halo) vs.\ disk luminosity contribution at
red wavelength (Gunn $r$ band). Kent's results substantially match also the 
near-infrared observations (see Fig.~\ref{st_calib}), while $B$ luminosity 
profiles \citep{simien86} tend in general to show a slightly
enhanced disk component in later-type spirals, as a consequence of a 
bluer color with respect to the bulge.

For each Hubble type we eventually calibrated a morphological parameter 
defined as S/T~$= L{\rm (spheroid)}/L{\rm (tot)}$. As the S/T calibration 
does not vary much at visual and infrared wavelength, we fixed the $I$ 
luminosity as a reference for model setup. This also fairly traces the 
bolometric partition (cf.\ the corresponding fitting coefficients of 
Table~\ref{coeff}), especially for early-type spirals.
According to the observations, the luminous mass of the disk remains roughly 
constant along the Sa--Im sequence, while bulge luminosity decreases towards 
later-type spirals \citep{simien83, gavazzi93}.

\subsection{The spheroid sub-system}

There is general consensus on the fact that both the halo and bulge sub-systems 
in the Galaxy basically fit with coeval SSPs older than $\sim 13$~Gyr 
\citep{gilmore,renzini93,frogel99,fg00}.
Observations of the central bulge of the Milky Way show that it mostly 
consists of metal-rich stars \citep{frogel88,frogel99} and this seems a quite 
common situation also for external galaxies \citep{jablonka,goudfrooij,davidge}.
The exact amount of bulge metallicity, however, has been subject to continual 
revision in the recent years, ranging from a marked metal overabundance 
(i.e.\ [Fe/H]~$\sim +0.2$; \citealp{wr83,rich,gf92}) to less prominent values, actually 
consistent with a standard  or even slightly sub-solar metallicity 
\citep{tiede,sadler,zoccali,origlia}.

On the contrary, the halo mostly consists of metal-poor stars \citep{zinn,sf87},
and its metallicity can be probed by means of the globular cluster 
environment. From the complete compilation of 149 Galactic globular 
clusters by \citet{harris}, for instance, we derive a mean [Fe/H]~$= -1.24 \pm 
0.56$, with clusters spanning a range $-2.3 \la$~[Fe/H]~$\la 0.0$. 
This figure is in line with the inferred metallicity distribution of globular 
cluster systems in external galaxies \citep{bh91,durrell,perrett,kp02}.

According to the previous arguments, for the spheroid component in our models
we will adopt two coeval SSPs with [Fe/H]$ = +0.22$ and $-1.24$, for bulge
and halo, respectively.\footnote{We chose to maintain a super-solar metallicity
for the bulge component, in better agreement with the observations of external 
galaxies.}
Once accounting for metallicity, via eq.~(\ref{eq:ml}) and Table~1, the standard halo/bulge 
mass ratio for the Milky Way \citep[e.g.][]{sandage87,dwek} translates 
into a relative bolometric luminosity
\begin{equation}
[L_{\rm halo} : L_{\rm bulge}] = [ 15\%: 85\%].
\end{equation}
This partition will be adopted throughout in our models and provides a nearly 
solar luminosity-weighted metallicity for the spheroid system as a whole.

\subsection{The disk}

To set the disk distinctive parameters we need to suitably constrain stellar 
birthrate $b$ (or, equivalently, the SFR power-law index, $\eta$), and mean 
stellar metallicity along the Hubble morphological sequence. 
This could be done relying on the observed colors of present-day galaxies. 
In our analysis we will assume a current age of 15 Gyr.

The most exhaustive collection of photometric data for local galaxies
definitely remains the RC3 catalog \citep{rc3}. Based on the
original database of over 2500 objects,  \citet{buta} carried out a 
systematic analysis of the optical color distribution.
Another comprehensive  compilation from the RC3-UGC catalogs (1537 galaxies 
in total) is that of \citet{rh94}.
Both data samples have been extensively discussed in Paper I; their analysis 
shows that a 1-$\sigma$ color scatter of the order of $\pm 0.15$ mag  can be
devised both for the $B-V$ and $U-B$ distributions as a realistic estimate of
the intrinsic spread within each $T$ morphology class \citep[cf.\ also][on this point]{fukugita95}.
This value should probably be increased by a factor of two 
for the infrared colors.

\begin{figure}
\centerline{
\psfig{file=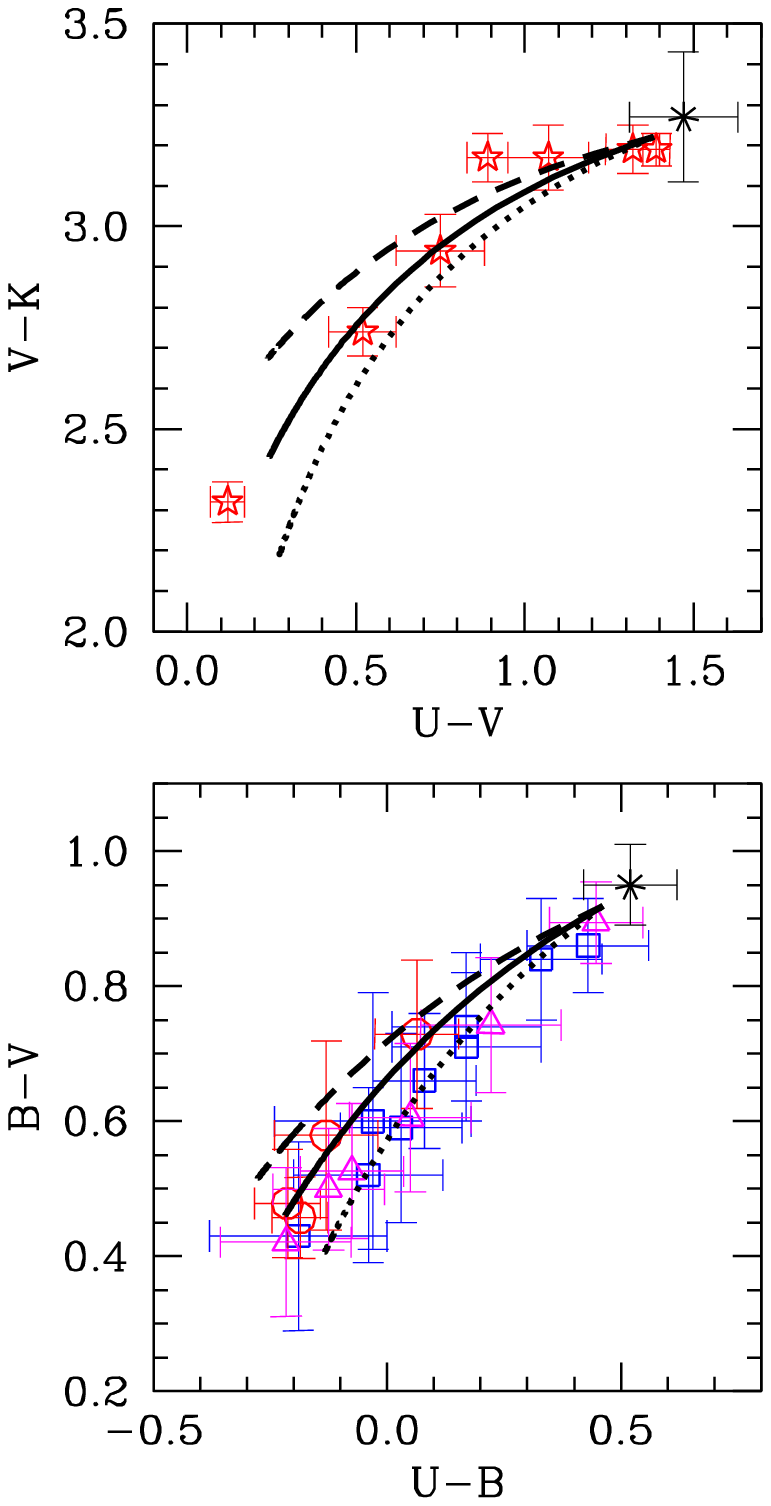,width=0.72\hsize,clip=}
}
\caption{Two-color diagrams of galaxy distribution compared with 15 Gyr
model sequences for different values of disk metallicity. Data are from 
\citet[][ $\star$ markers]{aaronson}, \citet[][$\circ$]{pence}, \citet[][$\sq$]{gavazzi91}, 
and \citet[][$\triangle$]{buta}. 
Mean colors for ellipticals (the $*$ marker in each panel) are from B95. 
In the models, a disk component is added as an increasing fraction of the 
total galaxy luminosity (in the sense of increasingly bluer colors). 
Long-dashed line is the theoretical locus for solar metallicity, while 
solid and dotted lines are for $[Fe/H]_{\rm disk} = -0.5$, and $-1.0$ dex, 
respectively.}
\label{col4}
\end{figure}

\subsubsection{Two-color diagrams and disk metallicity}

A two-color diagram is especially suitable to constrain disk metallicity. 
Our experiments show in fact that any change in the stellar birthrate will 
shift the integrated colors along the same mean locus, for a fixed value of 
[Fe/H].
In Fig.~\ref{col4}, a set of disk model sequences with varying metallicity 
is compared with the $U,B,V,K$ photometry from the works of \citet{pence}, 
\citet{aaronson}, \citet{gavazzi91}, and \citet{buta}. 
We only included those data samples with complete multicolor photometry 
avoiding to combine colors from different sources in the literature. 
The theoretical loci in the figure have been computed by adding to the same 
spheroid component an increasing fraction of disk luminosity, according to 
eq.~(\ref{eq:lgal}) and assuming a SFR with $\eta = -0.8$. Three values for
metallicity have been considered, namely [Fe/H]~$= 0.0, -0.5$, and --1.0 dex.

Both the $(U-V)$ vs.\ $(V-K)$ and $(U-B)$ vs.\ $(B-V)$ plots clearly point 
to a mean sub-solar metal content for the disk stellar component. This is 
especially constrained by late-type galaxies, where disk dominates total 
luminosity. We could tentatively adopt [Fe/H]$_{\rm disk} = -0.5$ dex as a 
luminosity-weighted representative value for our models.\footnote{In case of 
continual star formation, the luminosity-weighted  ``mean'' metallicity 
of a composite stellar population is in general lower than the actual [Fe/H] 
value of the youngest stars (and residual gas) in turn. This is because of 
the relative photometric contribution of the metal-poor unevolved component 
of low-mass stars, that bias the mean metal abundance toward lower values.\label{fehlum}}
As pointed out in Paper I, this value roughly agrees with the Milky Way stellar 
population in the solar neighborhood \citep{edvardsson}, and is 
in line with the \citet{aj91} theoretical estimates, 
suggesting a mean luminosity-weighted [Fe/H]$_{\rm disk} \sim -0.3$~dex for their 
disk-dominated galaxy models.
The intrinsic spread of the observations in Fig.~\ref{col4}, compared with 
the full range of the [Fe/H] loci, indicates however that metal abundance is 
not a leading parameter to modulate disk colors, and even a $\pm 0.3$~dex 
change in our assumptions would not seriously affect model predictions.

\begin{figure}
\centerline{
\psfig{file=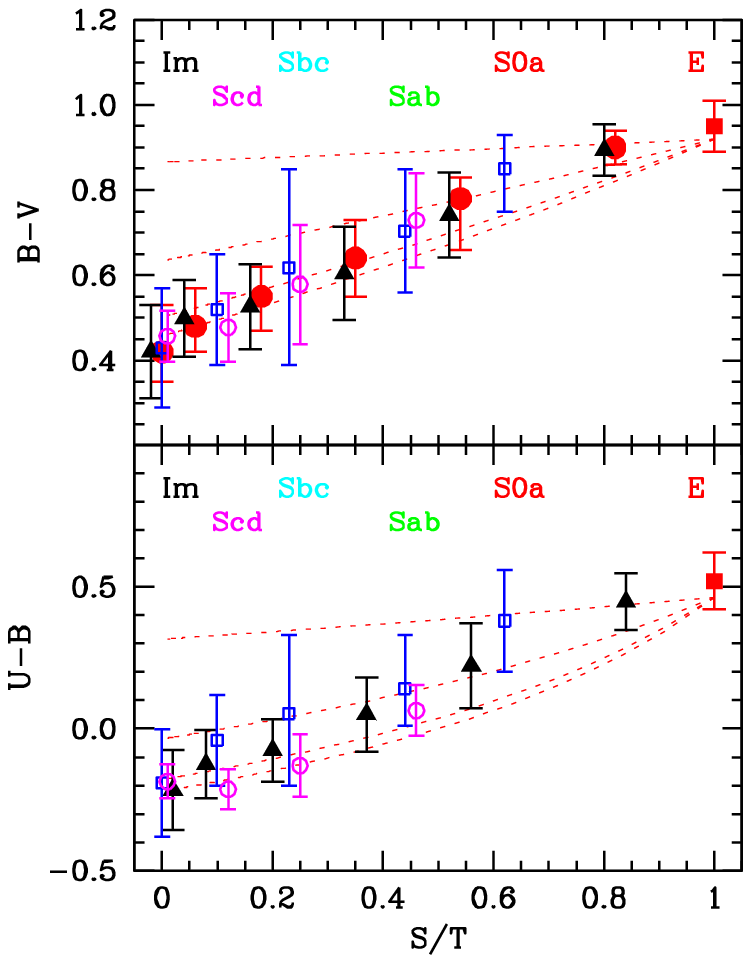,width=0.75\hsize,clip=}
}
\caption{Galaxy color distribution vs.\ bolometric morphological parameter 
S/T~$ = L_{\rm spheroid}/L_{\rm tot}$ compared with 15 Gyr model sequences 
with fixed SFR power-law index. Data are from \citet[][$\circ$ markers]{pence}, 
\citet[][$\sq$]{gavazzi91}, \citet[][$\bullet$]{rh94}, and \citet[][$\blacktriangle$]{buta}.
Mean colors for ellipticals ($\blacksquare$) are from B95. 
Dotted lines display the expected locus for models with three different values 
of the SFR index, namely $\eta = -0.8, 0$, and +0.8 plus the SSP case, 
in the sense of increasingly redder colors. A trend is evident in the 
observations with a higher stellar birthrate for later-type systems.}
\label{colst2}
\end{figure}

\subsubsection{Color distribution and disk SFR}

Color distribution along with the S/T morphological parameter is a useful tool
to constrain the disk SFR.  This plot is in fact better sensitive to the 
relative amount of young vs.\ old stars in the galaxy stellar population, 
and gives an implicit measure of the disk birthrate, $b$. Our results are 
summarized in Fig.~\ref{colst2}, comparing with the available $UBV$ photometry.

Four model sequences are displayed in each panel according to three different 
values of the SFR power-law index, namely $\eta = 0.8, 0.0, -0.8$ 
(i.e.\ $b = 0.2, 1, 1.8$) plus the case of a plain SSP evolution. We adopted 
[Fe/H]$_{\rm disk} = -0.5$~dex throughout in our calculations.
Both the $B-V$ and $U-B$ plots indicate, in average, a higher stellar birthrate
for later-type systems \citep{k94}. Our adopted calibration for 
$b$ vs.\ morphological type is shown in Fig.~8 of Paper I.

\begin{figure}
\centerline{
\psfig{file=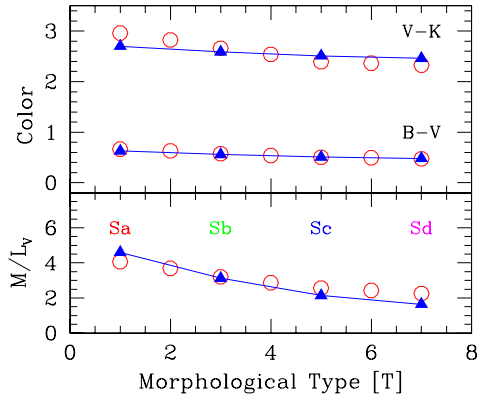,width=0.8\hsize,clip=}
}
\caption{A comparison of our disk model output ($\blacktriangle$ markers) 
vs.\ the \citet{aj91} original results ($\circ$ markers).
The $V$-band M/L ratio (in solar units) refers to the stellar mass alone 
(i.e. the total mass converted to stars at the age of 15 Gyr); it has been
normalized consistently, to account for the slightly different IMF mass range 
and slope (see text for discussion).}
\label{arimoto}
\end{figure}

\begin{table}
\caption{Input distinctive parameters for 15 Gyr models}
\begin{tabular}{ccrcc}
\hline
{Hubble Type} & {$b^{(a)}$} & {$\eta$} & {S/T$^{(b)}$}  & {[Fe/H]} \\
           &     &         &       & {[halo\ :\ disk\ :\ bulge]} \\
\hline
E\,~       & 0.0 & SSP     & 1.00 &  --1.24~~$\qquad$ ~~+0.22 \\
S0        & 0.0 & SSP     & 0.80 &  --1.24~~--0.50~~+0.22 \\
Sa        & 0.2 & $0.8$   & 0.55 &  --1.24~~--0.50~~+0.22 \\
Sb        & 0.5 & $0.5$   & 0.35 &  --1.24~~--0.50~~+0.22 \\
Sc        & 0.9 & $0.1$   & 0.18 &  --1.24~~--0.50~~+0.22 \\
Sd        & 1.3 & $-0.3$  & 0.05 &  --1.24~~--0.50~~+0.22 \\
Im        & 1.8 & $-0.8$  & 0.00 &  $\qquad$~~--0.50~~~$\qquad$ \\
\hline
\end{tabular}
\medskip

$^{(a)}${Disk stellar birthrate}\\
$^{(b)}${$S/T = L{\rm (spheroid)}/L{\rm (tot)}$ at red/infrared wavelength.}
\label{input}
\end{table}

It is useful to compare our final results with the models of \citet{aj91}, 
which addressed disk chemio-photometric evolution in deeper detail.
A slightly different IMF boundaries were adopted by \citet{aj91} 
(i.e.\ a power-law index $s = 2.45$ instead of our standard Salpeter value, 
$s = 2.35$, and a stellar mass range $M_* = [0.05,\ 60]~M_\odot$ vs.\ our 
value of $M_* = [0.1,\ 120]~M_\odot$), so that we had to rescale their 
original values to consistently compare with our M/L ratio.
Figure~\ref{arimoto} shows a remarkable agreement with our results, along the 
whole late-type galaxy sequence, both in terms of integrated colors and 
stellar M/L ratio. To account for the missing luminosity of 60-120~$M_\odot$ 
stars in the \citet{aj91} receipt, however, one should further 
decrease the inferred M/L ratio for their models in Fig.~\ref{arimoto}, 
especially for Sc-Sd systems.

\section {Model output}

The distinctive parameters eventually adopted for our galaxy templates, 
according to the previous discussion, have been collected in Table~\ref{input}.
A synoptic summary of the main output properties for 15 and 1~Gyr models is reported 
in Table~\ref{output}. Compared with Paper I models, notice that we slightly 
revised here the bolometric luminosity scale as a consequence of a refined 
fitting function in Table~\ref{coeff}. We therefore predict here a lower 
bolometric M/L ratio (because of a higher $L_{\rm bol}$ value), compared to Table~3 
of Paper I. Apart from this difference, the change has no effect on the rest 
of the model properties as bolometric luminosity is a {\it derived} quantity 
for our calculations; full consistency is therefore preserved with the 
Paper I framework.

\begin{figure}
\centerline{
\psfig{file=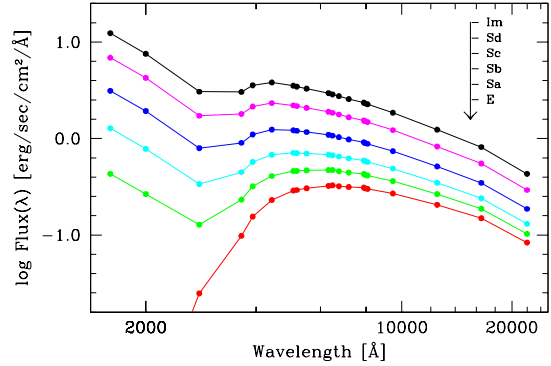,width=0.9\hsize,clip=}
}
\caption{Theoretical SED for 15 Gyr template models of $M_{gal} = 
10^{11}~M_\odot$ along the Hubble sequence. Multicolor magnitudes have been 
converted to monocromatic flux density according to the photometric zero 
points of Table~\ref{coeff}. The ultraviolet flux estimates at 1600, 2000, 
and 2800 \AA\ are from Paper I.}
\label{sed}
\end{figure}

Template galaxy models are described in full detail in the series of 
Table~\ref{tab_el} to \ref{tab_im} of Appendix A, assuming a total 
mass $M_{\rm gal} = 10^{11}$~M$_\odot$ for the system at 15 Gyr.\footnote{Throughout this paper, 
$M_{\rm gal}$ refers to the amount of mass converted to stars (that is, $M_{\rm gal} =
 \int{\rm SFR} dt$) and therefore it does {\it not} include residual gas. 
By definition, $M_{\rm gal}$ increases with time.\label{mgal}} 
In addition to the standard colors in the Johnson, Gunn and Washington photometric systems,
also ``composite'' colors like $(g-V)$ and $(M-V)$ are reported in the tables in order to
allow an easier transformation of photometry among the different systems.
It could also be useful to recall, in this regard, that a straightforward transformation
of the Johnson/Cousins magnitudes to the Sloan SDSS $(u',g',r',i',z')$ photometric 
system can be obtained relying on the equations set of \citet[][their eq.~23]{fukugita96}.
A plot of the synthetic SED 
for 15 Gyr galaxies is displayed in Fig.~\ref{sed} in a $\log$-$\log$ plot. 
Luminosity at different wavelength is obtained by converting theoretical 
magnitudes to absolute apparent flux, according to the photometric zero points
of Table~\ref{coeff} (column 3). In the figure, ultraviolet magnitudes at 
1600, 2000 and 2800 \AA\ are from Paper I.

\begin{table*}
\begin{minipage}{180mm}
\caption{Output summary for template galaxy models at 1 and 15 Gyr}
\begin{tabular}{cccccccccc}
\hline
\\
   & \multicolumn{4}{c}{$\mid$\hrulefill~~{\large 1 Gyr}~~\hrulefill$\mid$} &  \multicolumn{4}{c}{$\mid$\hrulefill~~{\large 15 Gyr}~~\hrulefill$\mid$} \\
\\
Hubble  & {S/T}$^{(a)}$  & {L/L$_{\rm tot}^{(b)}$} & {M/M$_{\rm tot}^{(c)}$} & {M/L}$^{(d)}$   &  {S/T}  & {L/L$_{\rm tot}$} & {M/M$_{\rm tot}$} & {M/L} \\
Type           &       & {[halo\ :\ disk\ :\ bulge]} & {[halo\ :\ disk\ :\ bulge]} &  &       & {[halo\ :\ disk\ :\ bulge]} & {[halo\ :\ disk\ :\ bulge]} &   \\
\hline
E\,~      & 1.00 & 0.15~~0.00~~0.85  & 0.09~~0.00~~0.91 & 0.74 &  1.00 & 0.15~~0.00~~0.85  & 0.09~~0.00~~0.91 & 6.41 \\
S0        & 0.81 & 0.12~~0.19~~0.69  & 0.08~~0.16~~0.76 & 0.71 &  0.81 & 0.12~~0.19~~0.69  & 0.08~~0.16~~0.76 & 6.14 \\
Sa        & 0.66 & 0.10~~0.34~~0.56  & 0.08~~0.17~~0.75 & 0.59 &  0.52 & 0.08~~0.48~~0.44  & 0.07~~0.26~~0.67 & 4.48 \\
Sb        & 0.62 & 0.09~~0.38~~0.53  & 0.07~~0.13~~0.80 & 0.53 &  0.30 & 0.04~~0.70~~0.26  & 0.05~~0.37~~0.58 & 3.06 \\
Sc        & 0.65 & 0.10~~0.35~~0.55  & 0.08~~0.08~~0.84 & 0.53 &  0.14 & 0.02~~0.86~~0.12  & 0.04~~0.51~~0.45 & 1.86 \\
Sd        & 0.56 & 0.08~~0.44~~0.48  & 0.08~~0.09~~0.83 & 0.47 &  0.04 & 0.01~~0.96~~0.03  & 0.02~~0.77~~0.21 & 1.05 \\
Im        & 0.00 & 0.00~~1.00~~0.00  & 0.00~~1.00~~0.00 & 0.07 &  0.00 & 0.00~~1.00~~0.00  & 0.00~~1.00~~0.00 & 0.66 \\
\hline
          & {[Fe/H]$_{\rm tot}^{(e)}$} & {Disk colors} & {Integrated colors}  &           & {[Fe/H]$_{\rm tot}$} & {Disk colors} & {Integrated colors}  &  \\
          &     &  {(U--V)~(B--V)~(V--K)} &  {(U--V)~(B--V)~(V--K)} &        &      &  {(U--V)~(B--V)~(V--K)} &  {(U--V)~(B--V)~(V--K)}     & \\ 
\hline
E\,~	  & $+0.00$ &			 & ~~0.74~~~0.66~~~2.82    & &  $+0.00$ &		     & 1.38~~~0.92~~~3.22 & \\
S0	  & $-0.09$ & ~~0.60~~~0.60~~~2.55 & ~~0.71~~~0.64~~~2.77  & &  $-0.09$ & 1.18~~~0.87~~~2.96 & 1.33~~~0.91~~~3.17 & \\
Sa	  & $-0.17$ & ~~0.20~~~0.43~~~2.34 & ~~0.53~~~0.58~~~2.68  & &  $-0.24$ & 0.60~~~0.63~~~2.70 & 0.91~~~0.76~~~2.98 & \\
Sb	  & $-0.19$ & ~~0.10~~~0.38~~~2.27 & ~~0.47~~~0.55~~~2.66  & &  $-0.34$ & 0.44~~~0.56~~~2.59 & 0.63~~~0.65~~~2.81 & \\
Sc	  & $-0.18$ & ~~0.01~~~0.34~~~2.20 & ~~0.45~~~0.55~~~2.66  & &  $-0.43$ & 0.34~~~0.51~~~2.51 & 0.43~~~0.56~~~2.63 & \\
Sd	  & $-0.21$ & --0.04~~~0.32~~~2.16 & ~~0.36~~~0.51~~~2.60  & &  $-0.49$ & 0.28~~~0.48~~~2.46 & 0.31~~~0.49~~~2.50 & \\
Im	  & $-0.50$ & --0.07~~~0.30~~~2.13 & --0.07~~~0.30~~~2.13  & &  $-0.50$ & 0.24~~~0.46~~~2.42 & 0.24~~~0.46~~~2.42 & \\
\hline

\end{tabular}
\smallskip

$^{(a)}${Bulge+ halo luminosity fraction in bolometric}\\
$^{(b)}${Luminosity partition in bolometric}\\
$^{(c)}${Mass partition of the stellar component}\\
$^{(d)}${Stellar mass-to-light ratio of the model, in bolometric and solar units.}\\
$^{(e)}${Luminosity-weighted mean metallicity of the model from the bolometric, defined as 
[Fe/H]$_{\rm tot} = \sum_j$ [Fe/H]$_j$ (L$_j$/L$_{\rm tot}$), with $j = 1,2,3$ for the three galaxy components 
(i.e.\ halo, disk and bulge). The adopted value of [Fe/H]$_j$ for each component is from Table~\ref{input}.}
\label{output}
\end{minipage}
\end{table*}

\begin{figure}
\centerline{
\psfig{file=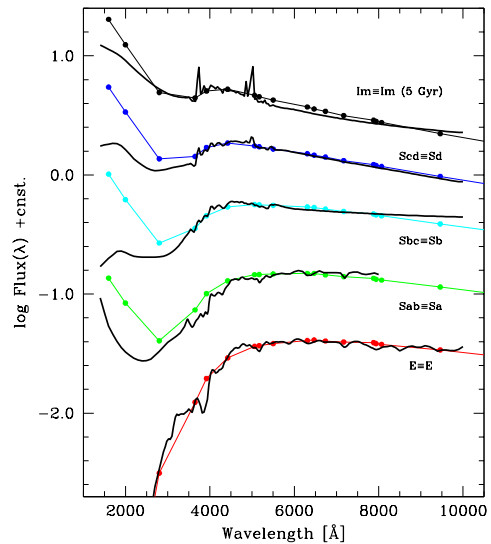,width=0.8\hsize,clip=}
}
\caption{The SED for template galaxy models (connected solid dots) is compared  with ``mean'' 
empirical spectra along the Hubble morphological sequence (thick solid line). Data 
for types E (the M81 case), Sbc, Scd, and Im are from \citet{coleman}, while
the Sab case is from \citet{pence}. Empirical and theoretical templates are matched
as labeled on each plot.
}
\label{coleman}
\end{figure}

A comparison of our output with the empirical SED for template galaxies
along the Hubble sequence is carried out in Fig.~\ref{coleman}. Mean reference spectra 
are those assembled by \citet{coleman} for types E (the M81 case), Sbc, Scd, and Im, while
the SED of the Sab type in the figure is from \citet{pence}. From the plots
one can appreciate a fully suitable match between observed and theoretical SED 
longward of the $U$ band. Two interesting features, however, are worth of attention, to
a deeper analysis: {\it i)} empirical templates (especially for spirals) always display a ``depressed'' 
ultraviolet emission, compared to the theoretical SED;  {\it ii)} the Im empirical template more closely
fits a young ($\sim 5$~Gyr) theoretical model.

As we have further discussed in Paper I, point {\it i)} is the obvious signature 
of dust in the SED of real galaxies. Although this effect is not
explicitely taken into account in our models, it could easily be assessed in any {\it ex-post}
analysis of the observations by adopting a preferred shape for the attenuation curve
to correct the data, like proposed, for example, by the studies of \citet{calzetti} 
and \citet{bmc88}.
In any case, as we were previously mentioning, it is evident from Fig.~\ref{coleman} that 
dust effects only enter at the very short wavelength range ($\lambda \la 3000$~\AA) of galaxy SED. 
Finally, as for point {\it ii)} above, one has to recall that the Coleman \etal reference 
template for the Im type relies on just one ``extremely bue irregular galaxy'' (i.e.\ NGC~4449). 
For this target, authors report, as integrated colors, $(U-V) = -0.06$ and $(B-V) = 0.32$, 
placing the object at the extreme blue range of the typical colors for irregulars (see, 
for instance, Fig.~\ref{col4} and \ref{colst2}, and also the relevant discussion 
by \citealp{fukugita95}); a younger age is therefore required to match the observed SED for 
this galaxy.

\begin{figure}
\centerline{
\psfig{file=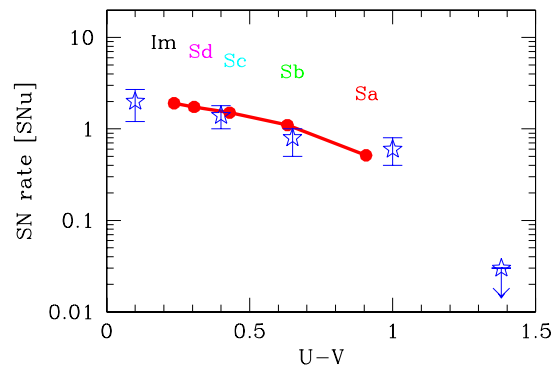,width=0.9\hsize,clip=}
}
\caption{Theoretical SN(II+Ibc) rates, singled out from eqs.~(\ref{eq:sn})
and (\ref{eq:snfraction}) (solid line 
and dots) are compared with the empirical estimates of \citet{cappellaro}
(star markers) from their survey of over 10\,000 low-redshift galaxies. 
The $U-V$ refers to the galaxy integrated color (corrected for reddening in 
case of observations).}
\label{sn_uv}
\end{figure}

\begin{figure*}
\centerline{
\psfig{file=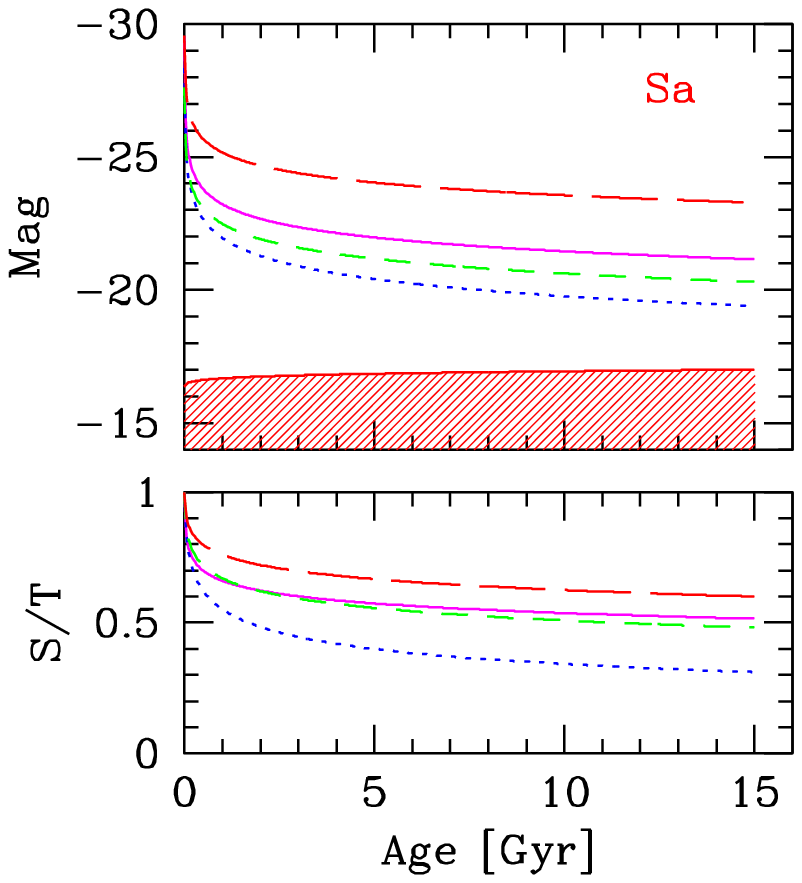,width=0.245\hsize,clip=}
\psfig{file=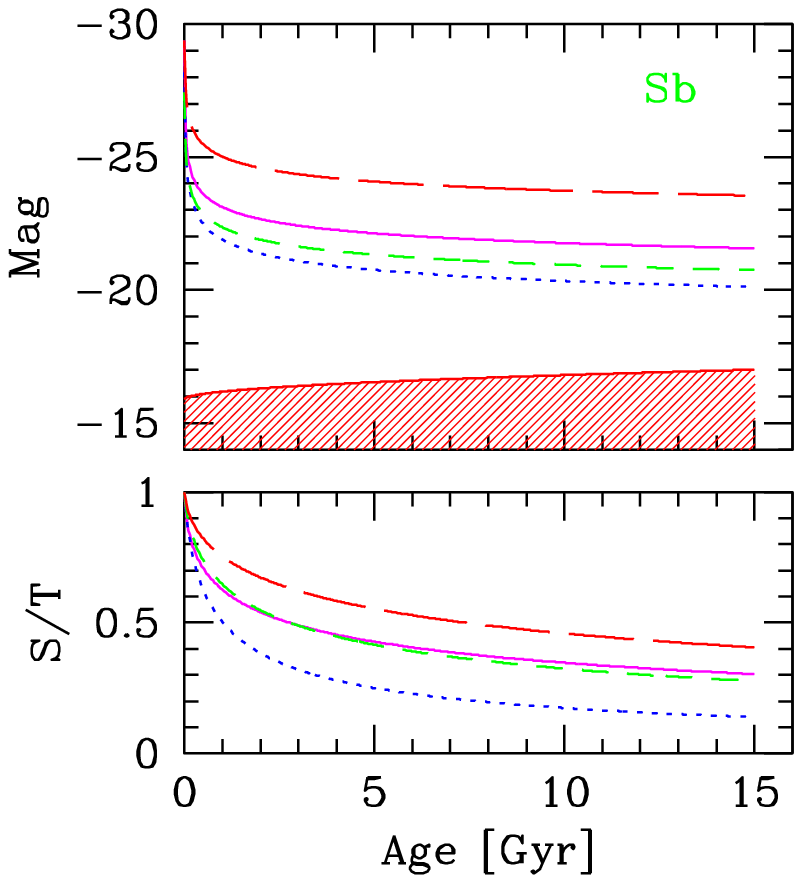,width=0.189\hsize,clip=}
\psfig{file=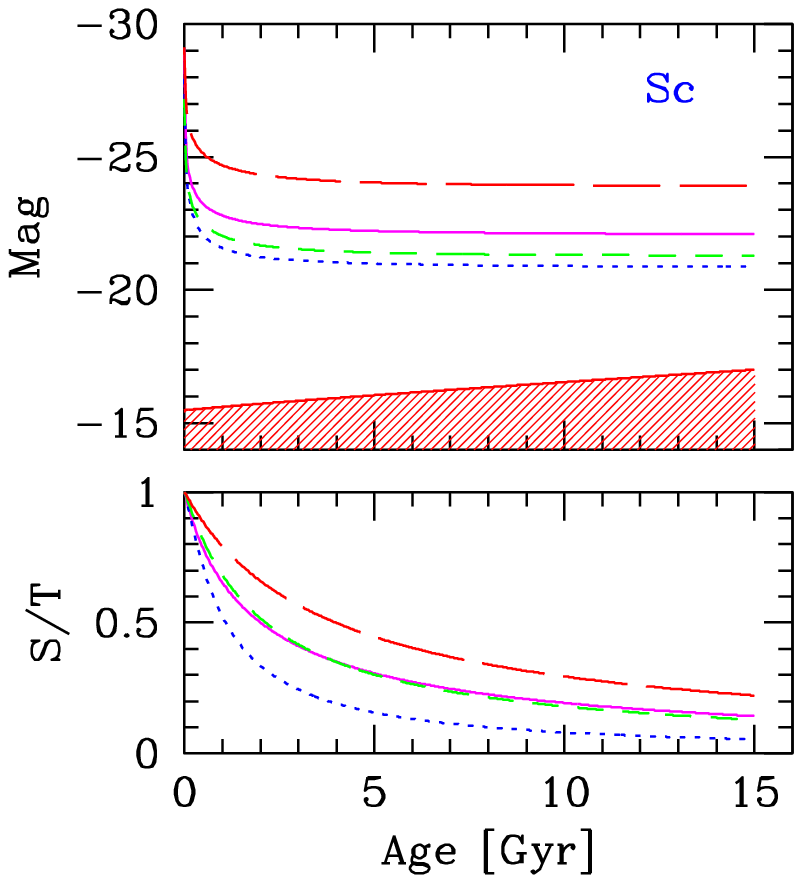,width=0.189\hsize,clip=}
\psfig{file=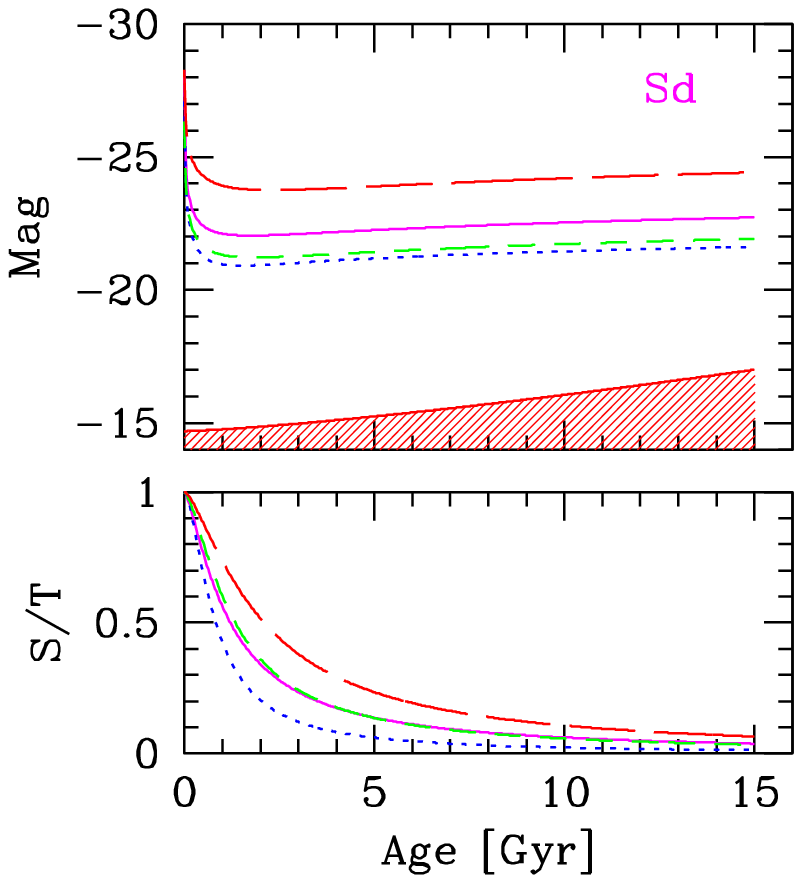,width=0.189\hsize,clip=}
\psfig{file=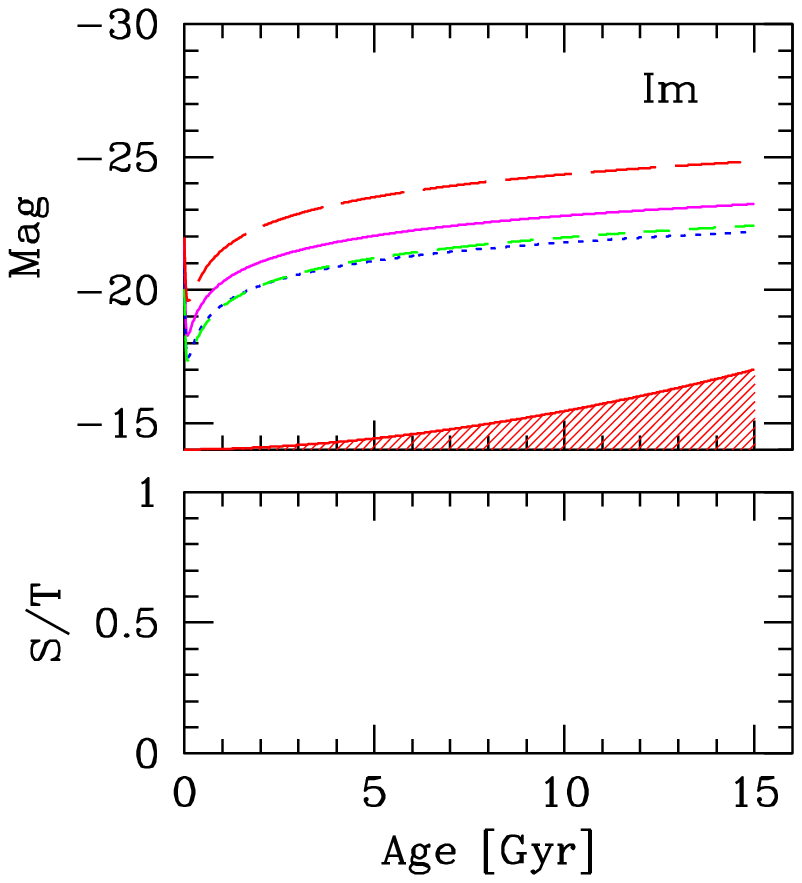,width=0.189\hsize,clip=}
}

\caption{{\it Upper plots in each panel:} theoretical luminosity evolution for 
late-type galaxy models.
A total stellar mass $M_{gal} = 10^{11}~M_\odot$ is assumed at 15 Gyr. Solid 
lines track bolometric magnitude, while dotted lines are for the $U$ band, 
short-dashed for $V$, and long-dashed for the $K$ band. The shaded area in each 
plot sketches (in arbitrary linear scale) the evolution of $M_{gal}$ according 
to star formation history for each morphological type.\protect \\
{\it Lower plots in each panel:} the corresponding evolution of the morhological
parameter S/T in the bolometric, $U, V$, and $K$ bands (same line caption as 
in the upper panels). Note that S/T~$\to 1$ at early epochs (excepting the 
Im model) due to the increasing bulge contribution.}
\label{tot}
\end{figure*}

\subsection {Supernova and Hypernova rates}

A natural output of template models deals with the current formation rate 
of core-collapsed stellar objects. This includes types II/Ibc Supernovae and 
their Hypernova (HN) variant \citep{paczynski} for very high-mass stars.
Both the SN and HN events are believed to generate from the explosion of 
single massive stars and have therefore a direct link with the galaxy stellar
birthrate. Hypernovae, in particular, have recently raised to a central 
issue in the investigation of the extragalactic Gamma-ray bursts 
\citep{nakamura, woosley}. 

If we assume, with \citet{bcf94}, that all stars with 
$M_* \ge 5~M_\odot$ eventually undergo an explosive stage, and those 
more massive than $40~M_\odot$ generate a Hypernova burst 
\citep{iwamoto}, then the number of SN(II+Ibc)+HN events in a SSP of 
total mass $M_{\rm SSP}$ (in solar units) and Salpeter IMF can be written as:
\begin{eqnarray}
N_{\rm SN+HN} &= &{0.35\over 1.35}\,\left[{{5^{-1.35}-120^{-1.35}}\over{0.1^{-0.35}-120^{-0.35}}}\right]\,M_{\rm SSP} \nonumber\\
              &= &0.0142\,M_{\rm SSP}\quad[M_\odot^{-1}].
\label{eq:snrate} 
\end{eqnarray}
The expected fraction of SN(II+Ibc) relative to HN candidates, for a Salpeter
IMF, derives as
\begin{equation}
[N_{\rm SN}: N_{\rm HN}]=[20:1].
\label{eq:snfraction}
\end{equation}

If disk SFR does not change much on a timescale comparable with lifetime of 
$5~M_\odot$ stars (i.e.\ $\sim 10^8$~yrs), then a simplified approach is allowed for 
composite stellar populations.
From eq.~(\ref{eq:birthrate}), at 15 Gyr we have that 
${\rm SFR}_o = b\, \langle{\rm SFR}\rangle = b\,{\rm M}_{\rm disk}/(15 \times 10^9 {\rm yr})$,
and the current SN+HN event rate, $R_{\rm SN+HN} =  dN_{\rm SN+HN}/dt$ can therefore be written as
\begin{equation}
R_{\rm SN+HN} = 0.0142~{\rm SFR}_o = 0.0142\,b\,f\,{{M_{\rm gal}}\over{(15 \times 10^9 {\rm yr})}},
\label{eq:snr}
\end{equation}
being $f$ the stellar mass fraction of disk (see Table~\ref{output}).
We could more suitably arrange eq.~(\ref{eq:snr}) in terms 
of bolometric mass-to-light ratio and total $B$ luminosity of the parent galaxy 
(again, cf.\ Table~\ref{output} for the reference quantities). 
With little arithmetic, $R_{\rm SN+HN}$ eventually becomes
\begin{equation}
R_{\rm SN+HN} = 0.95\,b\,f \left({M\over L_{\rm bol}}\right)\,10^{-0.4({\rm Bol - B} +0.69)}\,L_{10}^{\rm B}.
\label{eq:sn}
\end{equation}
In the equation, $L_{10}^{\rm B}$ is the galaxy $B$ luminosity in unit of 
$10^{10} L_{\odot}^{\rm B}$, while $(Bol-B)$ is the galaxy bolometric correction 
to the Johnson $B$ band, derived from Tables~\ref{tab_sa}-\ref{tab_im} 
as $(Bol-B) = (Bol -V) -(B-V)$.
With this notation, eq.~(\ref{eq:sn}) directly provides the SN+HN rate in the 
usual SNu units [i.e.\ 1\,SNu = 1\,SN(100yr)$^{-1}$($10^{10}L_{\odot}^{\rm B})^{-1}$].

\begin{table}
\caption{Theoretical SN\,(II+Ibc) and Hypernova rates for late-type galaxies at present time}
\begin{tabular}{ccccc}
\hline
Hubble & SN+HN$^{(a)}$ & HN$^{(a)}$ & $\tau_{SN+HN}^{(b)}$ & $\tau_{HN}^{(b)}$ \\
Type   & {[SNu]} &  {[SNu]} & {[yr]} & {[yr]} \\
\hline
Sa & 0.52 & 0.025 & 194  & 4100  \\
Sb & 1.10 & 0.052 & ~~91  & 1900  \\
Sc & 1.50 & 0.071 & ~~67  & 1400  \\
Sd & 1.74 & 0.082 & ~~58  & 1250  \\
Im & 1.91 & 0.090 & ~~52  & 1100  \\
\hline
\end{tabular}
\smallskip

$^{(a)}${SN(II+Ibc)+HN and HN rates, in SNu units.}\\
$^{(b)}${Expected timescale between two SN or HN events in a $10^{11}~M_\odot$ galaxy}
\label{sn}
\end{table}

Table~\ref{sn} summarizes our results for late-type galaxies. In addition 
to the theoretical event rate, we also reported in the table the typical
timescale elapsed between two SN and HN bursts in a $10^{11}~M_\odot$ galaxy.
As $R_{\rm SN+HN}$ tightly depends on SFR, it should correlate with galaxy 
ultraviolet colors; this is shown in Fig.~\ref{sn_uv}, where a nice agreement 
is found with the empirical SN rates from the recent low-redshift galaxy survey 
by \citet{cappellaro}.

\subsection {Back-in-time evolution}

Theoretical luminosity evolution for disk-dominated systems in the Johnson 
$U, V$, $K$ bands, and for the bolometric is displayed in the upper panels of 
Fig.~\ref{tot}; to each plot we also added a shaded curve that traces the value 
of $M_{\rm gal}$ vs.\ time.
Lower panels in the figure represent the expected evolution of the morphological 
parameter S/T for the same photometric bands.
One striking feature in the Sa-Sd plots is the increasing luminosity contribution 
of the bulge at early epochs ($L_{\rm bulge} \propto t^{-0.8}$ in bolometric, 
see Table~\ref{coeff}). This greatly compensates the drop in disk luminosity
($L_{\rm disk} \propto t\, L_{\rm bulge}$, from eq.~\ref{eq:lgal}, for a constant 
SFR), and acts in the sense of predicting more nucleated (S/T~$\to 1$) galaxies 
at high redshift compared with present-day (i.e.\ 15~Gyr) objects.

\begin{figure}
\centerline{
\psfig{file=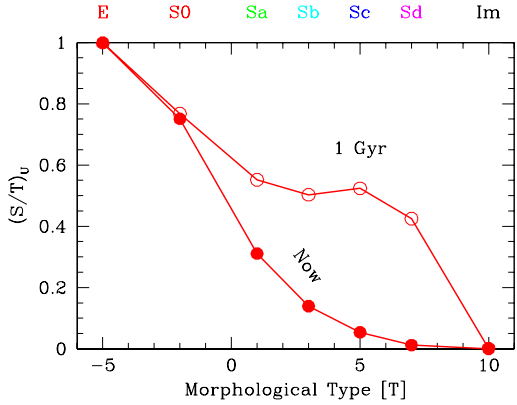,width=0.85\hsize,clip=}
}
\caption{Inferred evolution of the $U$-band morphological parameter S/T for 
galaxies along the Hubble sequence. Bulge enhancement at early epochs leads 
later-type systems (Sc-Sd types) at 1 Gyr to closely resemble present-day S0-Sa 
galaxies.}
\label{st_evol}
\end{figure}

This effect is shown in Fig.~\ref{st_evol}, where we track back-in-time 
evolution of the morphology parameter S/T in the ultraviolet range (Johnson $U$ 
band). Due to bulge enhancement, one sees that later-type spirals (Sc-Sd types)
at 1 Gyr closely resemble present-day S0-Sa systems.

\begin{figure}
\centerline{
\psfig{file=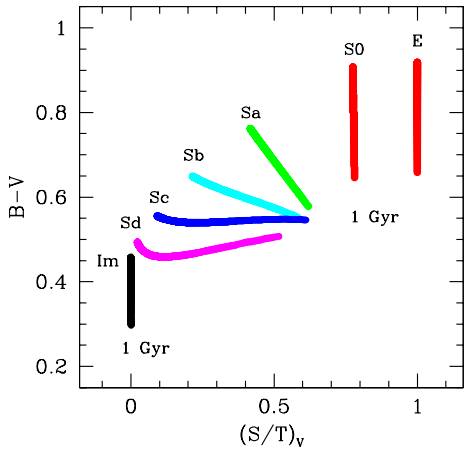,width=0.8\hsize,clip=}
}
\caption{Restframe $B-V$ color evolution, from 15 to 1 Gyr, vs.\ $V$-band 
morphological parameter S/T for galaxy models along the Hubble sequence. 
The increasing luminosity contribution of the bulge component back in time makes 
spiral galaxies at high redshift more nucleated than present-day homologues. 
For high-redshift observations this induces a bias toward both ``spiked'' and 
irregular systems with increasing distances (consider the {\it x}-axis 
projection of the figure) and conspires against the detection of grand-design 
spirals.}
\label{nuclei}
\end{figure}

A color vs.\ S/T plot, like in Fig.~\ref{nuclei}, effectively summarizes overall 
galaxy properties at the different ages.  Color evolution is much shallower for 
spirals than for ellipticals and, as expected, the trend is always in the sense 
of having bluer galaxies at earlier epochs (excepting perhaps Sd spirals) 
independently of the star formation details. Restframe colors 
tend however to ``degenerate'' with primeval late-type galaxies approaching 
the E model at early epochs as a consequence of the bulge brightening.

If one does not mind evolution, all these effects could lead to a strongly biased 
interpretation of high-redshift observations. For example, by relying on galaxy 
apparent colors (that is by reading Fig.~\ref{nuclei} from the {\it y}-axis, 
with no hints about morphology), the high-redshift galaxy population might 
show a lack of (intrinsically) red objects (ellipticals?), and enhance on the
contrary blue galaxies (spirals?). On the other hand, if we account for apparent
morphology alone (that 
is by reading Fig.~\ref{nuclei}  from the {\it x}-axis), then a bimodal excess 
of bulge-dominated systems (S/T~$\to 1$) and irregular star-forming galaxies 
(S/T~$\sim 0$) would appear at large distances \citep{vdb00, ky01}.
In fact, fiducial Im systems would eventually 
dominate with increasing redshift due to the disfavoring effect of k-correction 
on ellipticals, at least in the optical range. Among others, this should also 
conspire against the detection of grand-design spirals in high-redshift surveys
\citep[e.g.][]{vdb96}.

\begin{figure}
\centerline{
\psfig{file=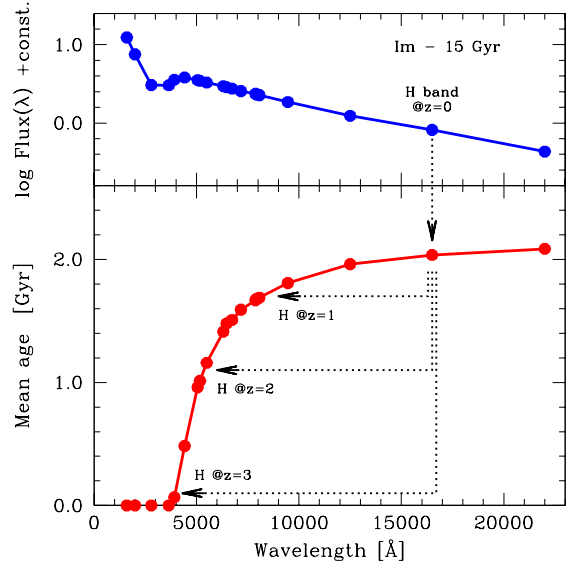,width=0.9\hsize,clip=}
}
\caption{Theoretical SED for the 15 Gyr Im galaxy model ({\it upper panel}), and ``mean'' age, 
$\overline{t}_*$, of the prevailing stars in the different photometric bands, according to eq.~(\ref{eq:ave2})
({\it lower panel}). Note that younger (and more massive) stars contribute, in average, 
to global luminosity at shorter wavelength. Assuming to observe this galaxy in the infrared H band
at increasing distances (namely, from $z = 0$ to 3, as labeled), one might be noticing 
an apparent increase in the fresh star-formation activity with redshift (and a correspondingly
younger inferred age) just as a consequence of probing blue/ultraviolet   
emission in the galaxy restframe.}
\label{tdisk}
\end{figure}

\subsection{Redshift and age bias}

The effect of redshift, and its induced selective sampling of galaxy stellar 
population with changing wavelength, has even more pervasive consequences, 
as far as we track evolution of star-forming systems at increasing distances.

According to eq.~(\ref{eq:lgal}), the mean luminosity-weighted age of stars 
contributing to galaxy luminosity can be written as 
\begin{equation}
\overline{t}_* = {{\int_0^t \tau\ L_{\rm SSP}(\tau) {\it SFR}(t-\tau)\ d\tau} \over {L_{\rm tot}(t)}}.
\label{eq:ave}
\end{equation}

The ``representative'' age of stars changes therefore across galaxy SED, since 
$L_{\rm SSP} \propto t^{-\alpha}$, and  $\alpha$ depends on wavelength 
(cf.\ Table~\ref{coeff}). As, in our parameterization, SFR~$\propto t^{-\eta}$, 
then eq.~(\ref{eq:ave}) becomes
\begin{equation}
\overline{t}_* = {{\int_0^t \tau^{1-\alpha}\ (t-\tau)^{-\eta}\ d\tau} \over {\int_0^t \tau^{-\alpha}\ (t-\tau)^{-\eta}\ d\tau}}
= {{1-\alpha}\over {2-\alpha-\eta}}\ t.
\label{eq:ave2}
\end{equation}

In bolometric, $\alpha \simeq 0.8$ that is, for a constant SFR, $\overline {t}_* \simeq 0.2\, t$:
at 15 Gyr, bright stars are therefore in average $\sim 3~$Gyr old.
As an instructive example, in Fig.~\ref{tdisk} we displayed the mean age of stars 
contributing to galaxy luminosity at different wavelength for our 15 Gyr Im 
template model. Note that the value of $\overline{t}_*$ smoothly decreases at 
shorter wavelength, reaching a cutoff about 4000~\AA, when $\alpha$  exceeds 
unity and $\overline {t}_*$ coincides with the lifetime of the highest-mass 
stars in the IMF (see Paper I for the important consequences of this feature on 
galaxy ultraviolet SED). As a consequence, when tracking redshift evolution of 
star-forming galaxies through a given optical/infrared photometric band, one would be 
left with the tricky effect that distant objects appear to be younger than 
local homologues {\it in spite of any intrinsic evolution}.

\section {Luminosity contribution from residual gas}

In addition to the prevailing role of stars, a little fraction of galaxy 
luminosity (especially in late-type systems) is provided also by residual gas.
Its contribution is both in terms of continuum emission, mainly from 
free-bound $e^-$ transitions in the H\,{\small II} regions, and emission-line enhancement.
This is the typical case of the Balmer series, for instance, but also some 
forbidden lines, like those of [O\,{\small II}] at 3727 \AA\ and [O\,{\small III}] 
at 5007 \AA, usually appear as a striking feature in the galaxy spectrum  
\citep{k92,sodre}.

The key triggering process for gas luminosity is the ultraviolet emission from 
short-living ($t \la 10^7$ yrs) stars of high mass ($M_* \ga 10~M_\odot$), 
that supply most of the ionizing Lyman photons in the H\,{\small II} regions. 
The presence of emission lines is, in this sense, the most direct probe of 
ongoing star formation in a galaxy.
If gas is optically thin and tracks the distribution of young stars, then 
one could expect a tight relationship between the actual  SFR (via the number 
of UV-emitting stars) and the strength of the Hydrogen emission lines.

For its complexity, a detailed treatment of the nebular emission is 
obviously beyond the scope of this paper 
(see, e.g., \citealp{stasinska} and \citealp{mbb03}, for a reference discussion on this 
subject). 
Here, we are rather interested in exploring the general trend of some relevant features,
like the Balmer emission lines, that could supply an effective tool for the
SFR diagnostics when compared with the observations.

\begin{figure}
\centerline{
\psfig{file=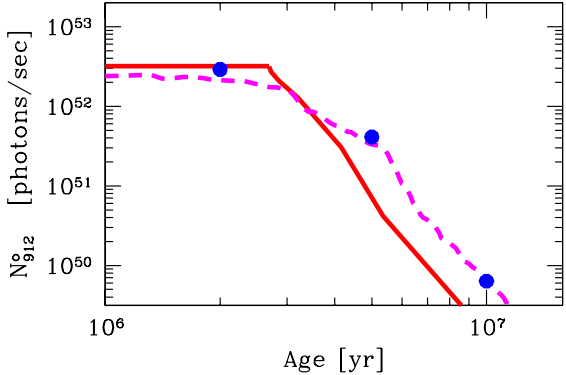,width=0.95\hsize,clip=}
}
\caption{The expected rate of Lyman photons, $N_{912}^o$, for a Salpeter SSP
of $10^6~M_\odot$ with stars in the range 0.1-120~$M_\odot$. Our results 
(thick solid line) consistently compare with the starburst models of 
\citet{leitherer} (dashed line), and \citet{mk91} 
(solid dots). Both these models have been rescaled to our output by matching 
the number of stars between 60 and $100~M_\odot$.}
\label{u912}
\end{figure}

\subsection {Nebular emission}

Our approach is similar to that of \citet{lh95}, adopting for 
the gas standard physical conditions, with $T = 10\,000$\,K and $Y = 0.28$ for 
Helium abundance (in mass). We will assume a full opacity to Lyman photons 
so that, once the photon rate $N_{912}^o$ from high-mass stars can be set from 
the detailed synthesis model, nebular continuum derives as
\begin{equation}
L_{gas}(\lambda) = {c\over{\lambda^2}} {\gamma_\lambda\over \alpha_B}~N_{912}^o.
\label{eq:nebular}
\end{equation}
In the equation, $\gamma(\lambda)$ is the continuum emission coefficient for 
the H-He chemical mix, according to \citet{aller}, and includes both 
free-free and bound-free transitions by Hydrogen and neutral Helium, as well 
as the two-photons continuum of Hydrogen. The Hydrogen recombination coefficient
(according to Case B of Baker and Menzel 1938) is from \citet{osterbrok} and 
has been set to $\alpha_B = 2.59~10^{-13}$\,cm$^3$~s$^{-1}$.

According to eq.~(\ref{eq:nebular}), the nebular continuum directly
scales with $N_{912}^o$ so that, as a relevant output of our models, in
Fig.~\ref{u912} we computed the expected rate of $\lambda < 912$~\AA\ 
photons for a Salpeter SSP with upper cutoff mass at 120~$M_\odot$.
Our results compare consistently with the starburst models of 
\citet{leitherer}, and  \citet{mk91}.
To account for different IMF slope and/or stellar mass limits, comparison
is made by matching our number of stars between 60 and $100~M_\odot$
in a $10^6\ M_\odot$ SSP. The agreement between the models is fairly good, 
with a tendency of our SSP, however, to evolve faster given a slightly shorter
lifetime assumed for high-mass stars (cf.\ Paper I for more details on 
the adopted stellar clock).

\begin{table*}
\begin{minipage}{140mm}
\caption{Balmer-line intensities for late-type galaxy templates$^{(a)}$}
\begin{tabular}{cccccccccccccccc}
\hline
  Age   & & \multicolumn{4}{c} {\hrulefill~~~Sa~~~\hrulefill} & & \multicolumn{4}{c} {\hrulefill~~Sb~~\hrulefill} & & \multicolumn{4}{c} {\hrulefill~~Sc~~\hrulefill} \\  
{[Gyr]} & & $H\delta$ &  $H\gamma$ &  $H\beta$ &  $H\alpha$ & &  $H\delta$ &  $H\gamma$ &  $H\beta$ &  $H\alpha$ & & $H\delta$ &  $H\gamma$ &  $H\beta$ &  $H\alpha$\\ 
\multicolumn{16}{c}{\hrulefill} \\
   1.0 & &  0.55 & 1.63 & 4.11 & 13.8 & & 0.79 & 2.33 & 6.01 & 20.6 & &  0.88 & 2.59 & 6.73 & 23.1 \\
   1.5 & &  0.56 & 1.63 & 4.05 & 13.3 & & 0.84 & 2.49 & 6.35 & 21.4 & &  1.03 & 3.02 & 7.86 & 26.9 \\
   2.0 & &  0.56 & 1.64 & 4.00 & 13.0 & & 0.88 & 2.58 & 6.55 & 21.9 & &  1.12 & 3.29 & 8.56 & 29.3 \\
   3.0 & &  0.56 & 1.64 & 3.93 & 12.4 & & 0.91 & 2.69 & 6.76 & 22.3 & &  1.22 & 3.58 & 9.36 & 32.1 \\
   4.0 & &  0.56 & 1.64 & 3.87 & 12.1 & & 0.93 & 2.75 & 6.87 & 22.5 & &  1.27 & 3.73 & 9.76 & 33.5 \\
   5.0 & &  0.56 & 1.64 & 3.83 & 11.8 & & 0.95 & 2.79 & 6.92 & 22.5 & &  1.29 & 3.80 & 9.97 & 34.2 \\
   6.0 & &  0.56 & 1.64 & 3.79 & 11.6 & & 0.95 & 2.81 & 6.95 & 22.5 & &  1.31 & 3.84 & 10.1 & 34.5 \\
   8.0 & &  0.55 & 1.63 & 3.73 & 11.2 & & 0.96 & 2.83 & 6.96 & 22.3 & &  1.31 & 3.87 & 10.1 & 34.6 \\
  10.0 & &  0.55 & 1.63 & 3.68 & 10.9 & & 0.97 & 2.84 & 6.95 & 22.1 & &  1.31 & 3.86 & 10.1 & 34.3 \\
  12.5 & &  0.55 & 1.62 & 3.62 & 10.6 & & 0.97 & 2.85 & 6.92 & 21.9 & &  1.30 & 3.84 & 9.98 & 33.9 \\
  15.0 & &  0.55 & 1.62 & 3.58 & 10.4 & & 0.97 & 2.84 & 6.88 & 21.6 & &  1.29 & 3.80 & 9.85 & 33.3 \\
\hline
\hline
  Age   & & \multicolumn{4}{c} {\hrulefill~~~Sd~~~\hrulefill} & & \multicolumn{4}{c} {\hrulefill~~Im~~\hrulefill} & & \multicolumn{4}{c} {~} \\  
{[Gyr]} & & $H\delta$ &  $H\gamma$ &  $H\beta$ &  $H\alpha$ & &  $H\delta$ &  $H\gamma$ &  $H\beta$ &  $H\alpha$ & &  \multicolumn{4}{c} {~} \\ 
\multicolumn{11}{c}{\hrulefill} & \multicolumn{5}{c}{~} \\
   1.0 &  & 1.20 & 3.54 & 9.46 & 33.4  & &  2.62 & 7.71 & 24.0 & 99.9  \\
   1.5 &  & 1.43 & 4.20 & 11.4 & 40.7  & &  2.42 & 7.13 & 21.9 & 90.1	\\
   2.0 &  & 1.54 & 4.54 & 12.5 & 45.0  & &  2.29 & 6.76 & 20.5 & 83.2	\\
   3.0 &  & 1.63 & 4.81 & 13.3 & 48.6  & &  2.13 & 6.28 & 18.7 & 74.6	\\
   4.0 &  & 1.65 & 4.87 & 13.5 & 49.5  & &  2.03 & 5.98 & 17.6 & 69.2	\\
   5.0 &  & 1.65 & 4.86 & 13.5 & 49.4  & &  1.95 & 5.76 & 16.8 & 65.3	\\
   6.0 &  & 1.64 & 4.82 & 13.4 & 48.8  & &  1.90 & 5.59 & 16.2 & 62.3	\\
   8.0 &  & 1.60 & 4.72 & 13.0 & 47.4  & &  1.81 & 5.34 & 15.3 & 57.9	\\
  10.0 &  & 1.57 & 4.64 & 12.7 & 45.9  & &  1.75 & 5.15 & 14.6 & 54.6	\\
  12.5 &  & 1.54 & 4.52 & 12.3 & 44.2  & &  1.69 & 4.97 & 14.0 & 51.6	\\
  15.0 &  & 1.50 & 4.43 & 12.0 & 42.7  & &  1.64 & 4.84 & 13.5 & 49.4	\\
\hline
\end{tabular}
\smallskip

$^{(a)}${The listed quantity is the Balmer emission-line equivalent width in \AA.}
\label{btable}
\end{minipage}
\end{table*}

\subsection{Balmer emission-line evolution}

The equivalent width of Balmer lines has been derived via the $H\beta$
luminosity, defined as
\begin{equation}
L(H\beta) = h\nu_{H\beta}\ {{\alpha_{H\beta}}\over {\alpha_B}}\ N_{912}^o.
\label{eq:hbeta}
\end{equation}
For the effective recombination coefficient at $T = 10\,000$~K we adopted 
the value $\alpha_{H\beta} = 3.03~10^{-14}$\,cm$^3$~s$^{-1}$ from \citet{osterbrok}. 
This eventually leads to a calibration for the $H\beta$ luminosity such as
\begin{equation}
L(H\beta) = 4.78~10^{-13}\ N_{912}^o~~~~[{\rm erg~s}^{-1}].
\label{eq:hbeta2}
\end{equation}
This result agrees within a 0.4\% both with the \citet{lh95} 
and \citet{copetti} calculations.
Balmer-line intensities, relative to $H\beta$, are from 
\citet[][Table 4.2 therein]{osterbrok} for the relevant value of the temperature.

If the continuum (including both the contribution from stars and gas, $L_*$
and $L_{\rm gas}$, respectively) is assumed to vary slowly with wavelength, 
adjacent to the line, then the $H\beta$ equivalent width can be written as
\begin{equation}
W(H\beta) = {{L(H\beta)}\over{(L_* + L_{gas})}}
\label{eq:whbeta}
\end{equation}
and all the other lines derive accordingly.
The computed value of $W(H\beta)$ should be regarded of course as the {\it net} 
emission from the gas component (that is after correcting the spectral feature
for stellar absorption).

\begin{figure*}
\centerline{
\psfig{file=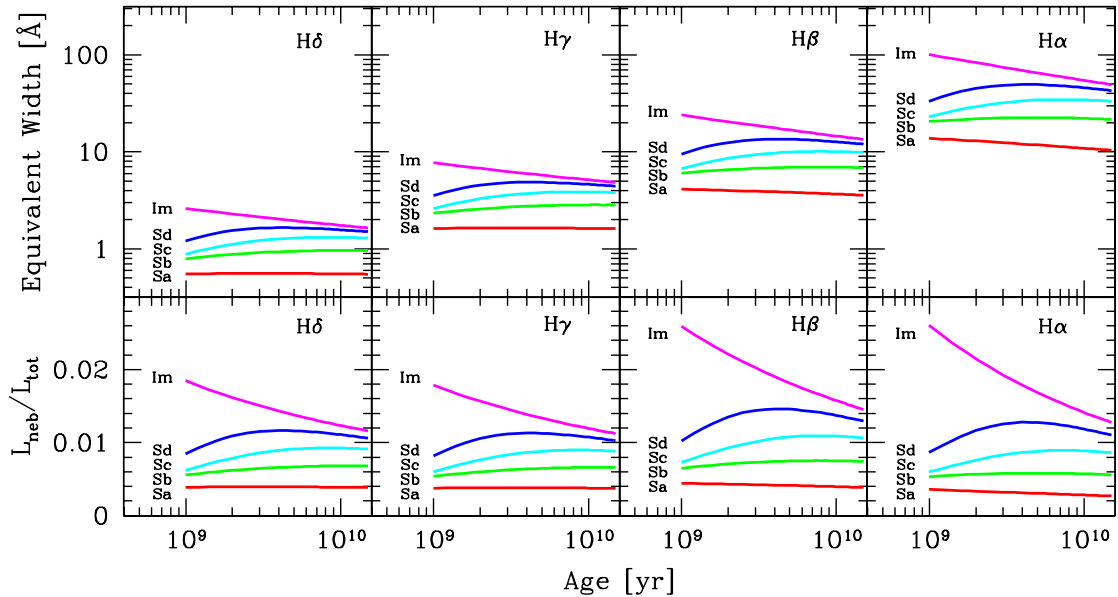,width=0.85\hsize,clip=}
}
\caption{Expected evolution for Balmer emission and nebular luminosity
from residual gas in our late-type galaxy templates.
Upper panel displays the equivalent width of each Balmer feature along 
galaxy age. Lower panel is the fraction of nebular luminosity supplied
to the total galaxy continuum evaluated at the relevant wavelength
close to each Balmer line.}
\label{balmer}
\end{figure*}

In Table~\ref{btable} we report our final results, also
summarized in Fig.~\ref{balmer}. The upper panel of the figure displays
the Balmer-line evolution for each of our template galaxy. As expected, 
$H\alpha$ is the dominating feature,
while gas emission sensibly decreases for $H\gamma$ and $H\delta$.
In addition, the onset of the galaxy bulge at early epochs works in the
sense of decreasing line emission because of the ``diluiting'' factor
$L_*$ in eq.~(\ref{eq:whbeta}) (cf.\ for example the diverging path of 
Im and Sd evolution in Fig.~\ref{balmer}).

In the lower panel of the same figure we computed the fraction 
$L_{\rm gas}/(L_* + L_{\rm gas})$ supplied by the nebular emission to the galaxy 
continuum at different wavelength, between 4100 and 6600 \AA, evaluated close 
to each Balmer line.
As far as the galaxy broad-band colors are concerned, we see that nebular
luminosity is almost negligible in our age range, and contributes at most by
a few percent to the galaxy total luminosity.

One striking feature that stems from the analysis of Fig.~\ref{balmer} is the
relative insensitivity of Balmer-line equivalent width to time. 
This is somewhat a consequence of the birthrate law assumed in our models. 
Actually, for the case of $H\beta$, like in eq.~(\ref{eq:whbeta}) for instance, we have that 
$L(H\beta)$ mainly responds to SFR$_o$, through the selective contribution of hot stars
of high mass, while the contiguous spectral continuum collects 
a much more composite piece of information from all stars (and it is, roughly, 
$L_* \propto \langle {\rm SFR} \rangle$); this makes $W(H\beta)$ better related to $b$.
In our framework, we could therefore conclude that observation of Balmer emissions
certainly provides important clues to size up the actual star formation activity in a galaxy, but
it will barely constrain age.

\section{Summary and conclusions}

In this work we attempted a comprehensive analysis of some relevant aspects of 
galaxy photometric evolution. Colors and basic morphological features for early- and late-type
systems have been reproduced by means of a set of theoretical population synthesis
models, that evaluate the individual photometric contribution of the three main stellar components
of a galaxy, namely the bulge, disk and halo.

Facing the formidable complexity of the problem (and the lack, to our present knowledge, 
of any straightforward ``prime principle'' governing the galaxy evolution), we chose to adopt a ``heuristic'' 
point of view, where the distinctive properties of present-day galaxies derive from a minimal 
set of physical assumptions and are mainly constrained by the observed colors along 
the Hubble sequence.

One important feature of our models is that galaxy SFR is a natural output of the gas-to-star 
conversion efficiency, that we assume to be an intrinsic and distinctive feature of galaxy 
morphological type (see eq.~\ref{eq:birthrate}). Our treatment of the star vs.~gas 
interplay is therefore somewhat different from other popular approaches, that rely on the
Schmidt law.\footnote{After all, even the Schmidt law is nothing more that one   
``reasonable'' (but nonetheless arbitrary) assumption in the current theory of 
late-type galaxy evolution. In this respect, for instance, there has been 
growing evidence that a classical dependence such as $\dot M_* \propto 
\rho_{gas}^n$ (whatever be the value for $n$) is to some extent inadequate 
for a general description of the SFR \citep{talbot,larson76,dopita,silk88,wyse,ryder}.}

Our results show that star formation history, as a function of the overall disk composition 
along the late-type galaxy sequence, appears to be a main factor to modulate galaxy 
photometric properties.
In general, the disk photometric contribution is the prevailing one in the 
luminosity (but not necessarily in the mass...) budget of present-day galaxies. Table~3 shows, for 
instance, that well less than a half of the total mass of a Sbc galaxy like the Milky Way is stored 
in the disk stars, while the latter provide over 3/4 of the global luminosity.

As shown in Fig.~\ref{colst2}, the observed colors of present-day galaxies tightly constrain 
the stellar birthrate leading to a smooth increasing trend for $b$ from E to Im types
(cf.\ Table~2 and also Fig.~8 in Paper I). 

The comparison with observed SN rate is an immediate ``acid test'' for our models, due to the marked 
sensitivity of this parameter to the current SFR.
The remarkably tuned match of our theoretical output with the SN observations 
in low-redshift galaxies (cf.\ Fig.~\ref{sn_uv}) is, in this regard, extremely encouraging.
As a possibly interesting feature for future observational feedback, 
we give in Table~4 also a prediction of the Hypernova event rate in late-type galaxies;
this quite new class of stars is raising an increasing interest for its claimed relationship
with the Gamma-ray bursts and the possible relevant impact of these events in the cosmological 
studies.

Another important point of our theoretical framework deals with the fact that galaxy evolution is
tracked in terms of the individual history of the different galaxy sub-systems.
This is a non-negligible aspect, as diverging evolutionary paths are envisaged 
for the bulge  vs.\ disk stellar populations. As discussed in Sec.~4.2, if 
$L_{\rm bulge} \propto t^{-0.8}$ in bolometric, and $L_{\rm disk} \propto t\,L_{\rm bulge}$, then
one has to expect that $L_{\rm bulge}/L_{\rm disk} \propto t^{-1}$,
that is the bulge always ends up as the dominant contributor to galaxy luminosity at early epochs.

As a consequence, the current morphological look of galaxies might  drastically change when
moving to larger distances, and we have shown in Sec.~4 how sensibly this bias could affect
the observation (and the interpretation) of high-redshift surveys.

In addition to broad-band colors, we have also briefly assessed the photometric contribution
of the nebular gas, studying in particular the expected evolution of Balmer line emission
in disk-dominated systems. As a main point in our analysis, models show that striking emission 
lines, like $H\alpha$, can very effectively track stellar birthrate in a galaxy. For these features to
be useful age tracers as well, however, one should first assess how $b$ could really change
with time on the basis of supplementary (and physically independent) arguments.

As a further follow-up of this work, we finally plan to complete the analysis of these galaxy 
template models providing, in a future paper, also the evolutionary $k$ corrections and other 
reference quantities for a wider application of the model output to
high-redshift studies.

\section*{Acknowledgments}

I wish to dedicate this work to my baby, Valentina, and to her mom
Claribel, for their infinite patience and invaluable support along the 
three years spent on this project.

\appendix

\section{Synthetic photometry for template galaxy models}
We report, in the series of Tables~\ref{tab_el}-\ref{tab_im}, the detailed output of the template
galaxy models for the different Hubble morphological types. All the models assume a total
{\it stellar} mass $M_{\rm gal} = (M_{\rm bulge} + M_{\rm disk} + M_{\rm halo}) = 10^{11}~M_\odot$ at 15 Gyr (see Footnote~\ref{mgal}
for an operational definition of $M_{\rm gal}$).

Full details on the photometric systems (i.e.\ band wavelength and magnitude zero points) can be 
obtained from Table~1.

Each table is arranged in two blocks of data; caption for entries in the upper block is the following:
\smallskip\\
Column 1 - Age of the galaxy, in Gyr;\\
Column 2 - Absolute magnitude of the model, in bolometric. For the Sun, we assume $Bol_\odot = 4.72$;\\
Column 3 - Bolometric correction to the Johnson $V$ band. The adopted solar value is $(Bol-V)_\odot = -0.07$;\\
Column 4 to 10 - Integrated broad-band colors in the Johnson system (filters $U,B,V,R,I,J,H,K$);
\smallskip\\
Lower block of data has the following entries:
\smallskip\\
Column 1 - Age of the galaxy, in Gyr;\\
Column 2 to 3 - Integrated colors in the Johnson/Cousins system (filters $R_c$ and $I_c$);\\
Column 4 to 6 - Integrated colors in the Gunn system (filters $g,r,i$), with the $g-V$ color
allowing a self-consistent link to the Johnson photometry;\\
Column 7 to 10 - Integrated colors in the Washington system (filters $C, M, T_1, T_2$), with
the $M-V$ color linking the Johnson photometry.
\smallskip

It may also be worth recalling that a straightforward transformation of our photometry to the Sloan SDSS photometric
system (extensively used in recent extragalactic studies) can be carried out according to the
equations set of \citet[][their eq.~23]{fukugita96}.

To allow an easier graphical display and interpolation of the data, all the magnitudes and colors 
in Tables~\ref{tab_el}-\ref{tab_im} are reported with a three-digit nominal precision; see, however, Sec.~2 and 4 for a
more detailed discussion of the real internal uncertainty of synthetic photometry in our models.

The entire theoretical database is publicly available at the author's Web site:
{\sf http://www.bo.astro.it/$\sim$eps/home.html}.

\begin{table*}
\begin{minipage}{100mm}
\caption{Template model for E galaxies}
\begin{tabular}{cccccccccc}
\hline
           &   & \multicolumn{8} {c}{\hrulefill~~Johnson~~\hrulefill} \\
{Age [Gyr]} &  {Bol} &  {Bol--V} &  {U--V} &  {B--V} &  {V--R} &  {V--I} &  {V--J} &  {V--H} & {V--K}\\ 
\hline
 1.0 & --23.100 & --0.722 &  0.740 &  0.656 &  0.698 &  1.301 &  1.916 &  2.630 &  2.821 \\
 1.5 & --22.736 & --0.745 &  0.840 &  0.696 &  0.725 &  1.343 &  1.968 &  2.688 &  2.882 \\
 2.0 & --22.493 & --0.764 &  0.907 &  0.723 &  0.744 &  1.371 &  2.004 &  2.727 &  2.923 \\
 3.0 & --22.142 & --0.792 &  1.003 &  0.763 &  0.771 &  1.413 &  2.055 &  2.783 &  2.983 \\
 4.0 & --21.888 & --0.814 &  1.073 &  0.791 &  0.791 &  1.443 &  2.093 &  2.824 &  3.027 \\
 5.0 & --21.698 & --0.832 &  1.125 &  0.813 &  0.806 &  1.465 &  2.121 &  2.855 &  3.059 \\
 6.0 & --21.543 & --0.847 &  1.168 &  0.830 &  0.818 &  1.484 &  2.144 &  2.880 &  3.086 \\
 8.0 & --21.298 & --0.870 &  1.235 &  0.858 &  0.837 &  1.513 &  2.181 &  2.920 &  3.128 \\
10.0 & --21.106 & --0.890 &  1.287 &  0.880 &  0.852 &  1.536 &  2.209 &  2.952 &  3.162 \\
12.5 & --20.919 & --0.909 &  1.338 &  0.902 &  0.867 &  1.559 &  2.237 &  2.983 &  3.194 \\
15.0 & --20.764 & --0.925 &  1.381 &  0.919 &  0.879 &  1.577 &  2.260 &  3.008 &  3.221 \\
\hline\hline
              & \multicolumn{2} {c}{\hrulefill~~Cousins~~\hrulefill} & \multicolumn{3} {c}{\hrulefill~~Gunn~~\hrulefill} & \multicolumn{4} {c}{\hrulefill~~Washington~~\hrulefill} \\
{Age [Gyr]} &  {V--R$_c$} &  {V--I$_c$} &  {g--V} &  {g--r} &  {g--i} &  {C--M} & {M--V} &  {M--T$_1$} &  {M--T$_2$} \\ 
\hline

 1.0 & 0.488 &  1.018 &  0.154 &  0.251 &  0.417     &  0.494 & 0.230 &  0.639 &  1.223  \\
 1.5 & 0.508 &  1.050 &  0.163 &  0.282 &  0.460     &  0.551 & 0.242 &  0.667 &  1.269  \\
 2.0 & 0.522 &  1.072 &  0.169 &  0.303 &  0.489     &  0.590 & 0.250 &  0.686 &  1.300  \\
 3.0 & 0.542 &  1.105 &  0.178 &  0.334 &  0.531     &  0.645 & 0.262 &  0.714 &  1.345  \\
 4.0 & 0.556 &  1.128 &  0.184 &  0.356 &  0.562     &  0.685 & 0.271 &  0.734 &  1.378  \\
 5.0 & 0.567 &  1.145 &  0.189 &  0.373 &  0.585     &  0.715 & 0.277 &  0.749 &  1.402  \\
 6.0 & 0.576 &  1.160 &  0.193 &  0.386 &  0.604     &  0.740 & 0.283 &  0.762 &  1.422  \\
 8.0 & 0.590 &  1.182 &  0.200 &  0.408 &  0.634     &  0.779 & 0.291 &  0.781 &  1.454  \\
10.0     & 0.601 &  1.200 &  0.204 &  0.425 &  0.658 &  0.809 & 0.298 &  0.797 &  1.479  \\
12.5     & 0.612 &  1.218 &  0.209 &  0.442 &  0.681 &  0.839 & 0.304 &  0.812 &  1.504  \\
15.0     & 0.621 &  1.232 &  0.213 &  0.456 &  0.700 &  0.864 & 0.310 &  0.824 &  1.524  \\
\hline
\end{tabular}
\label{tab_el}
\end{minipage}
\end{table*}

\begin{table*}
\begin{minipage}{100mm}
\caption{Template model for S0 galaxies}
\begin{tabular}{cccccccccc}
\hline
           &   & \multicolumn{8} {c}{\hrulefill~~Johnson~~\hrulefill} \\
 {Age [Gyr]} & {Bol} & {Bol--V} & {U--V} & {B--V} & {V--R} & {V--I} & {V--J} & {V--H} &{V--K}\\ 
\hline
  1.0 &  --23.146 & --0.699 &  0.710 &  0.644 &  0.686 &  1.277 &  1.881 &  2.584 &  2.769 \\
  1.5 &  --22.782 & --0.721 &  0.807 &  0.684 &  0.714 &  1.320 &  1.934 &  2.642 &  2.830 \\
  2.0 &  --22.539 & --0.738 &  0.872 &  0.711 &  0.733 &  1.348 &  1.969 &  2.681 &  2.871 \\
  3.0 &  --22.188 & --0.766 &  0.966 &  0.751 &  0.760 &  1.390 &  2.021 &  2.738 &  2.931 \\
  4.0 &  --21.934 & --0.787 &  1.034 &  0.779 &  0.780 &  1.420 &  2.058 &  2.779 &  2.975 \\
  5.0 &  --21.744 & --0.804 &  1.085 &  0.801 &  0.795 &  1.442 &  2.086 &  2.810 &  3.007 \\
  6.0 &  --21.589 & --0.818 &  1.126 &  0.818 &  0.807 &  1.461 &  2.109 &  2.835 &  3.034 \\
  8.0 &  --21.344 & --0.841 &  1.192 &  0.846 &  0.826 &  1.490 &  2.146 &  2.875 &  3.077 \\
  10.0 & --21.152 & --0.859 &  1.243 &  0.868 &  0.841 &  1.513 &  2.175 &  2.906 &  3.110 \\
  12.5 & --20.964 & --0.878 &  1.293 &  0.890 &  0.856 &  1.536 &  2.203 &  2.937 &  3.142 \\
  15.0 & --20.809 & --0.893 &  1.334 &  0.908 &  0.868 &  1.554 &  2.226 &  2.963 &  3.169 \\
\hline\hline
              & \multicolumn{2} {c}{\hrulefill~~Cousins~~\hrulefill} & \multicolumn{3} {c}{\hrulefill~~Gunn~~\hrulefill} & \multicolumn{4} {c}{\hrulefill~~Washington~~\hrulefill} \\
 {Age [Gyr]} & {V--R$_c$} & {V--I$_c$} & {g--V} & {g--r} & {g--i} & {C--M} & {M--V} & {M--T$_1$} & {M--T$_2$} \\ 
\hline
  1.0 &  0.480 &  1.000 &  0.152 &  0.239 &  0.398 &  0.471 &   0.226 &  0.629 &  1.204 \\
  1.5 &  0.500 &  1.033 &  0.161 &  0.270 &  0.441 &  0.526 &   0.239 &  0.657 &  1.250 \\
  2.0 &  0.514 &  1.055 &  0.167 &  0.291 &  0.470 &  0.564 &   0.247 &  0.676 &  1.281 \\
  3.0 &  0.534 &  1.087 &  0.176 &  0.322 &  0.513 &  0.618 &   0.259 &  0.704 &  1.326 \\
  4.0 &  0.548 &  1.110 &  0.182 &  0.344 &  0.543 &  0.657 &   0.268 &  0.724 &  1.359 \\
  5.0 &  0.559 &  1.128 &  0.187 &  0.361 &  0.567 &  0.686 &   0.274 &  0.740 &  1.383 \\
  6.0 &  0.568 &  1.142 &  0.191 &  0.375 &  0.586 &  0.711 &   0.279 &  0.752 &  1.403 \\
  8.0 &  0.582 &  1.165 &  0.197 &  0.397 &  0.616 &  0.749 &   0.288 &  0.772 &  1.435 \\
  10.0 & 0.593 &  1.182 &  0.202 &  0.414 &  0.639 &  0.778 &   0.295 &  0.787 &  1.460 \\
  12.5 & 0.604 &  1.200 &  0.207 &  0.430 &  0.662 &  0.808 &   0.301 &  0.802 &  1.485 \\
  15.0 & 0.613 &  1.214 &  0.211 &  0.444 &  0.682 &  0.832 &   0.307 &  0.815 &  1.505 \\
\hline
\end{tabular}
\label{tab_s0}
\end{minipage}
\end{table*}

\begin{table*}
\begin{minipage}{100mm}
\caption{Template model for S{\rm a} galaxies}
\begin{tabular}{cccccccccc}
\hline
           &   & \multicolumn{8} {c}{\hrulefill~~Johnson~~\hrulefill} \\
 {Age [Gyr]} &  {Bol} &  {Bol--V} &  {U--V} &  {B--V} &  {V--R} &  {V--I} &  {V--J} &  {V--H} & {V--K}\\ 
\hline
  1.0 &  --23.221 & --0.737 &  0.530 &  0.576 &  0.649 &  1.220 &  1.809 &  2.504 &  2.683 \\
  1.5 &  --22.895 & --0.750 &  0.594 &  0.605 &  0.671 &  1.253 &  1.850 &  2.548 &  2.730 \\
  2.0 &  --22.678 & --0.760 &  0.636 &  0.625 &  0.685 &  1.275 &  1.877 &  2.578 &  2.761 \\
  3.0 &  --22.365 & --0.776 &  0.695 &  0.653 &  0.706 &  1.307 &  1.917 &  2.620 &  2.806 \\
  4.0 &  --22.140 & --0.788 &  0.736 &  0.674 &  0.722 &  1.330 &  1.946 &  2.651 &  2.838 \\
  5.0 &  --21.972 & --0.798 &  0.766 &  0.689 &  0.733 &  1.348 &  1.967 &  2.674 &  2.862 \\
  6.0 &  --21.835 & --0.806 &  0.790 &  0.701 &  0.743 &  1.362 &  1.984 &  2.693 &  2.882 \\
  8.0 &  --21.620 & --0.820 &  0.827 &  0.720 &  0.757 &  1.384 &  2.012 &  2.723 &  2.913 \\
  10.0 & --21.452 & --0.831 &  0.856 &  0.735 &  0.769 &  1.402 &  2.034 &  2.746 &  2.938 \\
  12.5 & --21.288 & --0.841 &  0.884 &  0.750 &  0.780 &  1.419 &  2.055 &  2.769 &  2.962 \\
  15.0 & --21.153 & --0.850 &  0.907 &  0.762 &  0.790 &  1.433 &  2.073 &  2.788 &  2.982 \\
\hline\hline
              & \multicolumn{2} {c}{\hrulefill~~Cousins~~\hrulefill} & \multicolumn{3} {c}{\hrulefill~~Gunn~~\hrulefill} & \multicolumn{4} {c}{\hrulefill~~Washington~~\hrulefill} \\
 {Age [Gyr]} &  {V--R$_c$} &  {V--I$_c$} &  {g--V} &  {g--r} &  {g--i} & {C--M} & {M--V} &  {M--T$_1$} &  {M--T$_2$}  \\ 
\hline
  1.0  & 0.451 &  0.954 &  0.138 &  0.194 &  0.338 &  0.365  &   0.208 &  0.588 &  1.139 \\
  1.5  & 0.467 &  0.979 &  0.145 &  0.219 &  0.372 &  0.402  &   0.217 &  0.610 &  1.175 \\
  2.0  & 0.478 &  0.996 &  0.150 &  0.235 &  0.394 &  0.427  &   0.224 &  0.624 &  1.199 \\
  3.0  & 0.493 &  1.021 &  0.157 &  0.259 &  0.427 &  0.462  &   0.233 &  0.646 &  1.234 \\
  4.0  & 0.505 &  1.038 &  0.161 &  0.276 &  0.451 &  0.486  &   0.239 &  0.661 &  1.259 \\
  5.0  & 0.513 &  1.052 &  0.165 &  0.289 &  0.468 &  0.503  &   0.244 &  0.672 &  1.278 \\
  6.0  & 0.520 &  1.062 &  0.168 &  0.299 &  0.483 &  0.518  &   0.248 &  0.682 &  1.293 \\
  8.0  & 0.530 &  1.080 &  0.173 &  0.315 &  0.506 &  0.541  &   0.254 &  0.696 &  1.317 \\
  10.0 & 0.539 &  1.093 &  0.176 &  0.328 &  0.523 &  0.558  &   0.259 &  0.708 &  1.336 \\
  12.5 & 0.547 &  1.106 &  0.180 &  0.341 &  0.541 &  0.575  &   0.264 &  0.719 &  1.355 \\
  15.0 & 0.554 &  1.117 &  0.183 &  0.351 &  0.555 &  0.588  &   0.268 &  0.728 &  1.370 \\
\hline
\end{tabular}
\label{tab_sa}
\end{minipage}
\end{table*}

\begin{table*}
\begin{minipage}{100mm}
\caption{Template model for S{\rm b} galaxies}
\begin{tabular}{cccccccccc}
\hline
          &    & \multicolumn{8} {c}{\hrulefill~~Johnson~~\hrulefill} \\
 {Age [Gyr]} &  {Bol} &  {Bol--V} &  {U--V} &  {B--V} &  {V--R} &  {V--I} &  {V--J} &  {V--H} & {V--K}\\ 
\hline
  1.0 &  --23.114 & --0.757 &  0.467 &  0.551 &  0.635 &  1.199 &  1.784 &  2.477 &  2.655 \\
  1.5 &  --22.842 & --0.770 &  0.501 &  0.568 &  0.649 &  1.221 &  1.810 &  2.504 &  2.683 \\
  2.0 &  --22.667 & --0.778 &  0.521 &  0.579 &  0.658 &  1.234 &  1.826 &  2.520 &  2.700 \\
  3.0 &  --22.422 & --0.790 &  0.546 &  0.594 &  0.671 &  1.252 &  1.848 &  2.542 &  2.722 \\
  4.0 &  --22.252 & --0.797 &  0.562 &  0.604 &  0.679 &  1.265 &  1.863 &  2.557 &  2.737 \\
  5.0 &  --22.128 & --0.802 &  0.573 &  0.611 &  0.686 &  1.274 &  1.874 &  2.568 &  2.749 \\
  6.0 &  --22.030 & --0.806 &  0.583 &  0.617 &  0.691 &  1.282 &  1.883 &  2.577 &  2.758 \\
  8.0 &  --21.878 & --0.812 &  0.597 &  0.627 &  0.699 &  1.294 &  1.897 &  2.592 &  2.772 \\
  10.0 & --21.763 & --0.817 &  0.609 &  0.634 &  0.706 &  1.303 &  1.908 &  2.603 &  2.784 \\
  12.5 & --21.653 & --0.821 &  0.621 &  0.642 &  0.713 &  1.313 &  1.920 &  2.615 &  2.796 \\
  15.0 & --21.565 & --0.824 &  0.632 &  0.649 &  0.718 &  1.321 &  1.929 &  2.624 &  2.806 \\
\hline\hline
              & \multicolumn{2} {c}{\hrulefill~~Cousins~~\hrulefill} & \multicolumn{3} {c}{\hrulefill~~Gunn~~\hrulefill} & \multicolumn{4} {c}{\hrulefill~~Washington~~\hrulefill} \\
 {Age [Gyr]} &  {V--R$_c$} &  {V--I$_c$} &  {g--V} &  {g--r} &  {g--i} &  {C--M} & {M--V} &  {M--T$_1$} &  {M--T$_2$}  \\ 
\hline
  1.0  &  0.441 &  0.937 &  0.133 &  0.178 &  0.316 &  0.328  &   0.201 &  0.573 &  1.116 \\
  1.5  &  0.451 &  0.953 &  0.138 &  0.194 &  0.338 &  0.347  &   0.207 &  0.587 &  1.139 \\
  2.0  &  0.458 &  0.963 &  0.141 &  0.204 &  0.351 &  0.359  &   0.211 &  0.596 &  1.154 \\
  3.0  &  0.467 &  0.977 &  0.144 &  0.217 &  0.370 &  0.373  &   0.216 &  0.608 &  1.174 \\
  4.0  &  0.473 &  0.986 &  0.147 &  0.227 &  0.383 &  0.382  &   0.220 &  0.616 &  1.188 \\
  5.0  &  0.478 &  0.993 &  0.149 &  0.234 &  0.393 &  0.388  &   0.222 &  0.622 &  1.198 \\
  6.0  &  0.481 &  0.999 &  0.151 &  0.239 &  0.401 &  0.394  &   0.224 &  0.627 &  1.206 \\
  8.0  &  0.487 &  1.008 &  0.153 &  0.248 &  0.413 &  0.402  &   0.228 &  0.635 &  1.219 \\
  10.0 &  0.492 &  1.015 &  0.155 &  0.256 &  0.423 &  0.409  &   0.231 &  0.642 &  1.230 \\
  12.5 &  0.497 &  1.022 &  0.157 &  0.263 &  0.433 &  0.417  &   0.233 &  0.648 &  1.240 \\
  15.0 &  0.501 &  1.029 &  0.159 &  0.269 &  0.442 &  0.423  &   0.236 &  0.653 &  1.249 \\
\hline
\end{tabular}
\label{tab_sb}
\end{minipage}
\end{table*}

\begin{table*}
\begin{minipage}{100mm}
\caption{Template model for S{\rm c} galaxies}
\begin{tabular}{cccccccccc}
\hline
          &    & \multicolumn{8} {c}{\hrulefill~~Johnson~~\hrulefill} \\
 {Age [Gyr]} &  {Bol} &  {Bol--V} &  {U--V} &  {B--V} &  {V--R} &  {V--I} &  {V--J} &  {V--H} & {V--K}\\ 
\hline
  1.0 &  --22.799 & --0.770 &  0.453 &  0.546 &  0.633 &  1.199 &  1.786 &  2.480 &  2.659 \\
  1.5 &  --22.594 & --0.788 &  0.449 &  0.547 &  0.638 &  1.204 &  1.791 &  2.484 &  2.662 \\
  2.0 &  --22.478 & --0.798 &  0.441 &  0.546 &  0.639 &  1.205 &  1.790 &  2.481 &  2.658 \\
  3.0 &  --22.341 & --0.808 &  0.425 &  0.543 &  0.639 &  1.203 &  1.786 &  2.473 &  2.648 \\
  4.0 &  --22.264 & --0.812 &  0.416 &  0.541 &  0.639 &  1.201 &  1.782 &  2.466 &  2.639 \\
  5.0 &  --22.218 & --0.813 &  0.411 &  0.540 &  0.639 &  1.200 &  1.779 &  2.461 &  2.633 \\
  6.0 &  --22.187 & --0.813 &  0.409 &  0.540 &  0.639 &  1.200 &  1.778 &  2.458 &  2.629 \\
  8.0 &  --22.149 & --0.813 &  0.410 &  0.542 &  0.641 &  1.202 &  1.778 &  2.456 &  2.626 \\
10.0 & --22.128 & --0.812 &  0.415 &  0.546 &  0.644 &  1.205 &  1.781 &  2.457 &  2.626 \\
12.5 & --22.115 & --0.811 &  0.422 &  0.550 &  0.648 &  1.209 &  1.785 &  2.460 &  2.628 \\
15.0 & --22.108 & --0.810 &  0.430 &  0.555 &  0.651 &  1.214 &  1.791 &  2.465 &  2.633 \\
\hline\hline
              & \multicolumn{2} {c}{\hrulefill~~Cousins~~\hrulefill} & \multicolumn{3} {c}{\hrulefill~~Gunn~~\hrulefill} & \multicolumn{4} {c}{\hrulefill~~Washington~~\hrulefill} \\
 {Age [Gyr]} &  {V--R$_c$} &  {V--I$_c$} &  {g--V} &  {g--r} &  {g--i} &  {C--M} & {M--V} &  {M--T$_1$} &  {M--T$_2$}  \\ 
\hline
  1.0 & 0.440 &  0.936 &  0.132 &  0.176 &  0.314  &  0.321 &   0.200 &  0.570 &  1.114  \\
  1.5 & 0.443 &  0.940 &  0.133 &  0.180 &  0.320  &  0.317 &   0.201 &  0.574 &  1.120  \\
  2.0 & 0.443 &  0.940 &  0.134 &  0.181 &  0.321  &  0.311 &   0.202 &  0.574 &  1.121  \\
  3.0 & 0.443 &  0.938 &  0.134 &  0.180 &  0.320  &  0.300 &   0.201 &  0.574 &  1.120  \\
  4.0 & 0.443 &  0.936 &  0.134 &  0.180 &  0.319  &  0.294 &   0.201 &  0.573 &  1.119  \\
  5.0 & 0.443 &  0.935 &  0.134 &  0.180 &  0.319  &  0.290 &   0.201 &  0.573 &  1.119  \\
  6.0 & 0.443 &  0.935 &  0.134 &  0.181 &  0.319  &  0.288 &   0.202 &  0.574 &  1.119  \\
  8.0 & 0.445 &  0.936 &  0.135 &  0.183 &  0.322  &  0.288 &   0.203 &  0.576 &  1.122  \\
 10.0 &  0.446 &  0.938 &  0.136 &  0.186 &  0.326 &  0.290 &   0.204 &  0.579 &  1.126  \\
 12.5 &  0.449 &  0.942 &  0.137 &  0.190 &  0.331 &  0.294 &   0.205 &  0.582 &  1.132  \\
 15.0 &  0.452 &  0.946 &  0.138 &  0.194 &  0.336 &  0.299 &   0.207 &  0.586 &  1.138  \\
\hline
\end{tabular}
\label{tab_sc}
\end{minipage}
\end{table*}

\begin{table*}
\begin{minipage}{100mm}
\caption{Template model for S{\rm d} galaxies}
\begin{tabular}{cccccccccc}
\hline
          &    & \multicolumn{8} {c}{\hrulefill~~Johnson~~\hrulefill} \\
 {Age [Gyr]} &  {Bol} &  {Bol--V} &  {U--V} &  {B--V} &  {V--R} &  {V--I} &  {V--J} &  {V--H} & {V--K}\\ 
\hline
  1.0 & --22.117 & --0.796 &  0.362 &  0.508 &  0.610 &  1.163 &  1.739 &  2.428 &  2.603 \\
  1.5 & --22.050 & --0.818 &  0.314 &  0.489 &  0.600 &  1.144 &  1.713 &  2.395 &  2.566 \\
  2.0 & --22.049 & --0.826 &  0.283 &  0.476 &  0.591 &  1.129 &  1.692 &  2.368 &  2.536 \\
  3.0 & --22.105 & --0.829 &  0.252 &  0.463 &  0.582 &  1.110 &  1.665 &  2.332 &  2.496 \\
  4.0 & --22.180 & --0.827 &  0.242 &  0.459 &  0.578 &  1.103 &  1.652 &  2.315 &  2.475 \\
  5.0 & --22.252 & --0.823 &  0.242 &  0.459 &  0.578 &  1.101 &  1.648 &  2.308 &  2.466 \\
  6.0 & --22.319 & --0.820 &  0.246 &  0.462 &  0.579 &  1.102 &  1.648 &  2.306 &  2.463 \\
  8.0 & --22.436 & --0.814 &  0.259 &  0.468 &  0.584 &  1.107 &  1.653 &  2.309 &  2.465 \\
10.0 & --22.536 & --0.810 &  0.273 &  0.476 &  0.589 &  1.115 &  1.661 &  2.317 &  2.473 \\
12.5 & --22.640 & --0.806 &  0.290 &  0.485 &  0.596 &  1.124 &  1.673 &  2.329 &  2.485 \\
15.0 & --22.728 & --0.804 &  0.306 &  0.493 &  0.602 &  1.134 &  1.684 &  2.341 &  2.497 \\
\hline\hline
              & \multicolumn{2} {c}{\hrulefill~~Cousins~~\hrulefill} & \multicolumn{3} {c}{\hrulefill~~Gunn~~\hrulefill} & \multicolumn{4} {c}{\hrulefill~~Washington~~\hrulefill} \\
 {Age [Gyr]} &  {V--R$_c$} &  {V--I$_c$} &  {g--V} &  {g--r} &  {g--i} &  {C--M} &  {M--V} &  {M--T$_1$} &  {M--T$_2$}   \\ 
\hline
  1.0 & 0.422 &  0.908 &  0.124 &  0.149 &  0.277    &   0.266 & 0.189 &  0.546 &  1.074  \\
  1.5 & 0.414 &  0.892 &  0.121 &  0.137 &  0.259    &   0.235 & 0.184 &  0.534 &  1.056  \\
  2.0 & 0.408 &  0.880 &  0.119 &  0.127 &  0.245    &   0.213 & 0.181 &  0.526 &  1.041  \\
  3.0 & 0.401 &  0.866 &  0.116 &  0.116 &  0.229    &   0.193 & 0.177 &  0.517 &  1.024  \\
  4.0 & 0.399 &  0.860 &  0.115 &  0.113 &  0.223    &   0.186 & 0.176 &  0.514 &  1.018  \\
  5.0 & 0.399 &  0.859 &  0.115 &  0.113 &  0.222    &   0.185 & 0.177 &  0.514 &  1.017  \\
  6.0 & 0.399 &  0.859 &  0.116 &  0.114 &  0.224    &   0.187 & 0.177 &  0.515 &  1.019  \\
  8.0 & 0.403 &  0.864 &  0.117 &  0.119 &  0.231    &   0.194 & 0.179 &  0.520 &  1.027  \\
10.0     & 0.407 &  0.870 &  0.119 &  0.125 &  0.239 &   0.202 & 0.182 &  0.525 &  1.035  \\
12.5     & 0.412 &  0.877 &  0.121 &  0.133 &  0.249 &   0.212 & 0.185 &  0.532 &  1.046  \\
15.0     & 0.416 &  0.884 &  0.123 &  0.140 &  0.259 &   0.222 & 0.188 &  0.538 &  1.057  \\
\hline
\end{tabular}
\label{tab_sd}
\end{minipage}
\end{table*}

\begin{table*}
\begin{minipage}{100mm}
\caption{Template model for I{\rm m} galaxies}
\begin{tabular}{cccccccccc}
\hline
           &   & \multicolumn{8} {c}{\hrulefill~~Johnson~~\hrulefill} \\
 {Age [Gyr]} &  {Bol} &  {Bol--V} &  {U--V} &  {B--V} &  {V--R} &  {V--I} &  {V--J} &  {V--H} & {V--K}\\ 
\hline
  1.0 & --20.298 & --0.929 & --0.074 &  0.299 &  0.450 &  0.895 &  1.379 &  2.001 &  2.134 \\
  1.5 & --20.743 & --0.901 & --0.030 &  0.321 &  0.467 &  0.920 &  1.412 &  2.036 &  2.171 \\
  2.0 & --21.043 & --0.883 & ~\,0.001 &  0.337 &  0.479 &  0.939 &  1.435 &  2.062 &  2.199 \\
  3.0 & --21.481 & --0.861 & ~\,0.046 &  0.360 &  0.497 &  0.967 &  1.470 &  2.101 &  2.240 \\
  4.0 & --21.800 & --0.847 & ~\,0.080 &  0.377 &  0.510 &  0.988 &  1.497 &  2.131 &  2.272 \\
  5.0 & --22.040 & --0.838 & ~\,0.106 &  0.390 &  0.520 &  1.004 &  1.518 &  2.153 &  2.296 \\
  6.0 & --22.236 & --0.832 & ~\,0.127 &  0.401 &  0.529 &  1.018 &  1.534 &  2.172 &  2.316 \\
  8.0 & --22.547 & --0.822 & ~\,0.161 &  0.419 &  0.543 &  1.039 &  1.561 &  2.202 &  2.348 \\
10.0  & --22.791 & --0.816 & ~\,0.188 &  0.432 &  0.553 &  1.056 &  1.583 &  2.226 &  2.373 \\
12.5  & --23.030 & --0.811 & ~\,0.214 &  0.446 &  0.564 &  1.073 &  1.604 &  2.249 &  2.398 \\
15.0  & --23.229 & --0.807 & ~\,0.236 &  0.457 &  0.573 &  1.087 &  1.622 &  2.269 &  2.419 \\
\hline\hline
              & \multicolumn{2} {c}{\hrulefill~~Cousins~~\hrulefill} & \multicolumn{3} {c}{\hrulefill~~Gunn~~\hrulefill} & \multicolumn{4} {c}{\hrulefill~~Washington~~\hrulefill} \\
 {Age [Gyr]} &  {V--R$_c$} &  {V--I$_c$} &  {g--V} &  {g--r} &  {g--i} &  {C--M} & {M--V} &  {M--T$_1$} &  {M--T$_2$}  \\ 
\hline
  1.0 & 0.306 &  0.702 &  0.077 & --0.028 &  0.022  & --0.012 &   0.125 &  0.390 &  0.807  \\
  1.5 & 0.318 &  0.722 &  0.082 & --0.009 &  0.048  & ~\,0.016 &   0.132 &  0.406 &  0.834  \\
  2.0 & 0.327 &  0.736 &  0.086 & ~\,0.004 &  0.067 & ~\,0.037 &   0.137 &  0.418 &  0.854  \\
  3.0 & 0.340 &  0.757 &  0.092 & ~\,0.024 &  0.095 & ~\,0.064 &   0.145 &  0.436 &  0.883  \\
  4.0 & 0.350 &  0.773 &  0.096 & ~\,0.039 &  0.116 & ~\,0.084 &   0.150 &  0.449 &  0.906  \\
  5.0 & 0.357 &  0.786 &  0.099 & ~\,0.050 &  0.132 & ~\,0.100 &   0.154 &  0.459 &  0.922  \\
  6.0 & 0.363 &  0.796 &  0.101 & ~\,0.060 &  0.145 & ~\,0.113 &   0.158 &  0.467 &  0.936  \\
  8.0 & 0.373 &  0.812 &  0.106 & ~\,0.075 &  0.166 & ~\,0.134 &   0.163 &  0.481 &  0.959  \\
10.0  & 0.381 &  0.825 &  0.109 & ~\,0.086 &  0.183 & ~\,0.150 &   0.168 &  0.491 &  0.976  \\
12.5  & 0.389 &  0.838 &  0.112 & ~\,0.098 &  0.200 & ~\,0.166 &   0.172 &  0.502 &  0.994  \\
15.0  & 0.395 &  0.848 &  0.115 & ~\,0.108 &  0.213 & ~\,0.179 &   0.176 &  0.510 &  1.009  \\
\hline
\end{tabular}
\label{tab_im}
\end{minipage}
\end{table*}

\bsp
\label{lastpage}

\end{document}